\documentclass[onecolumn,showpacs,preprintnumbers,amsmath,amssymb]{revtex4}

\usepackage{graphicx}
\usepackage{dcolumn}
\usepackage{bm}

\begin{document}

\preprint{APS/123-QED}

\title{Lattice Boltzmann Model for High-Order Nonlinear Partial Differential Equations}

\author{Baochang Shi}
\email{shibc@hust.edu.cn}
\affiliation{%
School of Mathematics and Statistics, \\ Huazhong University of
Science and Technology, Wuhan 430074, People's Republic of China
}%

\author{Nanzhong He}
\email{nzhe@hust.edu.cn}
\affiliation{%
School of Mathematics and Statistics, \\ Huazhong University of
Science and Technology, Wuhan 430074, People's Republic of China
}%

\author{Zhaoli Guo}
\email{zlguo@hust.edu.cn} \affiliation{State Key Laboratory of Coal
Combustion, \\ Huazhong University of Science and Technology, Wuhan
430074, People's Republic of China
}%

\date{\today}

\begin{abstract}
A general lattice Boltzmann (LB) model is proposed for solving
nonlinear partial differential equations with the form $\partial_t
\phi+\sum_{k=1}^{m} \alpha_k \partial_x^k \Pi_k (\phi)=0$, where
$\alpha_k$ are constant coefficients, and $\Pi_k (\phi)$ are the
known differential functions of $\phi$, $1\leq k\leq m \leq 6$. The
model can be applied to the common nonlinear evolutionary equations,
such as (m)KdV equation, KdV-Burgers equation, K($m,n$) equation,
Kuramoto-Sivashinsky equation, and Kawahara equation, etc. Unlike
the existing LB models, the correct constraints on moments of
equilibrium distribution function in the proposed model are given by
choosing suitable \emph{auxiliary-moments},
and how to exactly recover the macroscopic equations through Chapman-Enskog expansion
 is discussed in this paper. Detailed simulations of
these equations are performed, and it is found that the numerical
results agree well with the analytical solutions and the numerical
solutions reported in previous studies.
\end{abstract}

\maketitle

\section{\label{sec:level1}Introduction}


The lattice Boltzmann method (LBM) is a promising technique for
simulating fluid flows and modeling complex physics in fluids
\cite{bsv,qso,cd}. Compared with the conventional computational
fluid dynamics approaches, the LBM is easy for programming,
intrinsically parallel, and it is also easy to include complicated
boundary conditions such as those in porous media. Up to now, the
most widely used LBM is the so-called lattice Bhatnagar-Gross-Krook
(LBGK) model. However, the LBGK model may suffer from numerical
instability when it is used to simulate the fluid with small
viscosity. A lot of work has been done to improve the stability of
LB model, among which the multi-relaxation-time LBM
\cite{Higuera,humieres,ll} and entropic LBM
\cite{Karlin,Ansumali,Chikatamarla,Boghosian,Keating} have attracted
much attention in recent years. It should be noted that the LBM also
shows potentials to simulate the nonlinear systems, such as the
reaction-diffusion equation \cite{dcd,bs,ys1}, convection-diffusion
equation (CDE) \cite{sooy,gsw,hlg,sddc}, Burgers equation
\cite{ys2}, KdV-like equation \cite{ChaiShiZheng}, Poisson equation
\cite{ChaiShi}, etc. Recently, the LB models have been extended to
solve CDEs on rectangular or irregular lattices \cite{se,sman} and
anisotropic dispersion equations \cite{zbc,rsm,g}, among which the
model proposed by Ginzburg \cite{g} is generic.

Except for solving real-valued nonlinear systems, the LB and LB-like
models have been successful in solving complex-valued nonlinear
systems. Since the middle of 1990s, several types of quantum lattice
gases and quantum LBM have been proposed based on quantum-computing
ideas to model some real and complex mathematical-physical
equations, such as the Schr\"{o}dinger equation, Gross-Pitaevskii
equation, Burgers equation, KdV equation
\cite{m,sb,succi,bt,yb,yepez,vyv,vvy,ps1,ps2}, etc. We refer the
readers to a recent paper \cite{ps2} for a detailed review. On the
other hand, recently the classical LB model has been used to model
complex-valued equations. In Ref. \cite{zfdg} the LBM was applied to
one-dimensional (1D) nonlinear Schr\"{o}dinger equation (NLSE)
following the idea of quantum lattice-gas model \cite{bt,yb} to
treat the reaction term. In Ref. \cite{shi}, motivated by the work
in Ref. \cite{zfdg}, the LBM for $n$-dimensional ($n$D) CDE with a
source term was directly applied to some nonlinear complex
equations, including the NLSE, coupled NLSEs, Klein-Gordon equation
and coupled Klein-Gordon-Schr\"{o}dinger equations, by adopting a
complex-valued distribution function and relaxation time. In Ref.
\cite{ShiGuo}, a general LB model for a class of $n$D nonlinear CDEs
was presented by properly selecting equilibrium distribution
function. The model in Ref.  \cite{ShiGuo} can be applied to both
real and complex-valued nonlinear evolutionary equations. Following
the idea in Ref. \cite{ShiGuo}, a LB model for 1D nonlinear Dirac
equation was given in Ref. \cite{ShiGuo2}, which is of second-order
accuracy in both space and time, and the order of accuracy is near
3.0 at lower grid resolution. The studies in Refs.
\cite{zfdg,shi,ShiGuo,ShiGuo2} show that the LBM may be an effective
numerical solver for real and complex-valued nonlinear systems.

Most of the existing LB models are used for solving partial
differential equations (PDEs) with order lower than or equal to
three, while many efficient conventional numerical methods for
solving higher-order PDEs have been proposed, such as the
pseudo-spectral method \cite{Lopez}, local discontinuous Galerkin
(LDG) method \cite{XuShu,XuShuRev}, finite different scheme
\cite{Ceballos}, finite volume method \cite{Cueto}, radial basis
function method \cite{Uddin}, among them LDG method has attracted
much attention due to its good nature, such as flexibility and high
parallel efficiency (see a recent review article \cite{XuShuRev} and
references therein for details). Although some higher-order LB
schemes have been proposed, they are mainly limited to solve
lower-order PDEs \cite{YanGW,CamasTsai,MaCF1} and 1D special
problems \cite{MaCF2}. Furthermore, the efficient numerical analysis
of these schemes are still needed. Therefore, it is important to
research and develop the LB model for solving higher-order PDEs.

In this paper, by extending the idea in
Ref. \cite{ShiGuo} a general LBGK model is proposed for solving a class of
nonlinear partial differential equations (NPDEs) with order up to six. In order to exactly
recover the macroscopic NPDE, the correct constraints on moments of
equilibrium distribution function in the proposed model are given by
introducing suitable \emph{auxiliary-moments}. The proposed model can be
used to solve many common nonlinear evolutionary equations,
such as (m)KdV equation, KdV-Burgers equation, K($m,n$) equation,
Kuramoto-Sivashinsky equation, and Kawahara equation, etc.
Numerical results show that the LBGK model can also be used to simulate higher-order NPDEs.

The rest of the paper is organized as follows. In Sec. II, the LBGK
model for NPDE is presented, and how to exactly recover the macroscopic equation from the model
 is discussed. In Sec. III, the equilibrium distribution and
auxiliary-moment functions for NPDEs with different orders are
given. Numerical tests of the LBGK model are made in Sec. IV, and
finally a brief summary is given in Sec. V.

\section{Multi-Scale Lattice Boltzmann Equations}

\subsection{LBGK Model}

The 1D NPDE considered in this paper can be written as
\begin{equation}
\partial_t \phi+\sum_{k=1}^{m} \alpha_k \partial_x^k \Pi_k (\phi)=0,
\label{NPDE}
\end{equation}
where $\phi$ is a scalar function of position {\it x} and time {\it
t}, $\alpha_1=1$, $\alpha_k$ are constant coefficients, and $\Pi_k
(\phi)$ are known differential functions of $\phi$, $1\leq k\leq m
\leq 6$.

Our LBGK model is based on the D1Q\emph{b} lattice \cite{qso} with
\emph{b} velocity directions in 1D space. The evolution equation of
the distribution function in the model reads
\begin{equation}
f_j(x+c_j \Delta t,t+\Delta t)= f_j(x,t)
             -\frac{1}{\tau}[f_j(x,t)-f_j^{\mathrm{eq}}(x,t)],
\label{LBE}
\end{equation} where
$\{c_j,j=0,\ldots,b-1\}\subseteq\{0,c,-c,2c,-2c,3c,-3c\}$ is the set
of discrete velocity directions, $c=\Delta x/\Delta t$, $\Delta x$
and $\Delta t$ are lattice spacing and time step, respectively,
$\tau$ is the dimensionless relaxation time, and
$f_j^{\mathrm{eq}}(x,t)$ is the equilibrium distribution function
(EDF).

To solve Eq. (\ref{NPDE}) using Eq. (\ref{LBE}), we must give
appropriate EDF $f_j^{\mathrm{eq}}(x,t)$. By applying the idea in
Ref. \cite{ShiGuo} to the higher-order NPDE (\ref{NPDE}), the
following constrains on $f_j^{\mathrm{eq}}$ are given
\begin{eqnarray}
\sum_j f_j=\sum_j f_j^{\mathrm{eq}}=\phi,\sum_j c_j
f_j^{\mathrm{eq}}=\Pi_1,\nonumber\\ \sum_j c_j^k
f_j^{\mathrm{eq}}=\Pi_{k0}+\beta_k\Pi_k,k=2,\ldots,m \label{moments}
\end{eqnarray}
where $\beta_k$ are parameters, and $\Pi_{k0}$ are
\emph{auxiliary-moment} (AM) functions for correctly recovering Eq.
(\ref{NPDE}), which are determined later, $k=2,\ldots,m$.

\subsection{Multi-Scale Lattice Boltzmann Equations}

To derive the macroscopic equation (\ref{NPDE}), the Chapman-Enskog
(C-E) expansion in time and space is applied:
\begin{equation}
f_j=\sum_{k=0}^6 \epsilon^k f_j^{(k)},\partial_t=\sum_{k=1}^6
\epsilon^k\partial_{t_{k}},\partial_x=\epsilon\partial_{x_1},
\label{CE}
\end{equation}
where $\epsilon$ is a small expansion parameter. Using the first
formula in Eq. (\ref{moments}) and Eq. (\ref{CE}), we have
\begin{equation}
\sum_j f_j^{(k)}=0, k \geq 1. \label{Neq}
\end{equation}

By applying Taylor expansion to Eq. (\ref{LBE}), we get
\begin{eqnarray}
D_j f_j + \frac{\Delta t}{2}D_j^2 f_j + \cdots +\frac{\Delta
t^5}{720}D_j^6 f_j+\ldots = -\frac{1}{\tau \Delta t} (f_j -
f_j^{\texttt{eq}}), \label{Taylor}
\end{eqnarray}
where $D_j=\partial_t+c_j\partial_x$. Denote
$D_{1j}=\partial_{t_1}+c_j\partial_{x_1}$. Similar to Ref.
\cite{YanGW}, substituting Eq. (\ref{CE}) into Eq. (\ref{Taylor})
and treating the terms in order of $\epsilon^k$ separately gives
\begin{eqnarray}
f_j^{(0)}=f_j^{\mathrm{eq}},\\
D_{1j} f_j^{(0)} = -\frac{1}{\tau \Delta t}f_j^{(1)},\label{Oeps1}\\
\partial_{t_2} f_j^{(0)} + \tau_2 \Delta t D_{1j}^2 f_j^{(0)}= -\frac{1}{\tau \Delta t} f_j^{(2)},\label{Oeps2}\\
\partial_{t_3} f_j^{(0)} + 2\tau_2 \Delta t\partial_{t_2} D_{1j} f_j^{(0)}+\tau_3\Delta t^2D_{1j}^3 f_j^{(0)}= -\frac{1}{\tau \Delta t}
f_j^{(3)},\label{Oeps3}\\
\partial_{t_4} f_j^{(0)} + 2\tau_2 \Delta t\partial_{t_3} D_{1j} f_j^{(0)}+ 3\tau_3 \Delta t^2\partial_{t_2} D_{1j}^2 f_j^{(0)}
+\tau_2\Delta t \partial_{t_2}^2 f_j^{(0)}+\tau_4\Delta t^3D_{1j}^4 f_j^{(0)}= -\frac{1}{\tau \Delta t} f_j^{(4)},\label{Oeps4}\\
\partial_{t_5} f_j^{(0)} + 2\tau_2 \Delta t\partial_{t_4} D_{1j} f_j^{(0)}+ 3\tau_3 \Delta t^2\partial_{t_2}^2 D_{1j}
f_j^{(0)}+3\tau_3 \Delta t^2\partial_{t_3} D_{1j}^2
f_j^{(0)}\nonumber\\+ 2\tau_2 \Delta t\partial_{t_2} \partial_{t_3}
f_j^{(0)}+ 4\tau_4\Delta t^3\partial_{t_2} D_{1j}^3 f_j^{(0)}+
\tau_5\Delta t^4D_{1j}^5 f_j^{(0)}= -\frac{1}{\tau \Delta t} f_j^{(5)},\label{Oeps5}\\
\partial_{t_6} f_j^{(0)} + 2\tau_2 \Delta t\partial_{t_5} D_{1j} f_j^{(0)}+ 6\tau_3 \Delta
t^2\partial_{t_2}\partial_{t_3}D_{1j} f_j^{(0)}+3\tau_3 \Delta
t^2\partial_{t_4} D_{1j}^2 f_j^{(0)}+ 2\tau_2 \Delta t\partial_{t_2}
\partial_{t_4} f_j^{(0)}+ 6\tau_4\Delta t^3\partial_{t_2}^2 D_{1j}^2
f_j^{(0)} \nonumber\\ +\tau_3 \Delta t^2\partial_{t_2}^3 f_j^{(0)}
+\tau_2 \Delta t\partial_{t_3}^2 f_j^{(0)}+4\tau_4\Delta
t^3\partial_{t_3} D_{1j}^3 f_j^{(0)}+5\tau_5\Delta t^4\partial_{t_2}
D_{1j}^4 f_j^{(0)}+\tau_6\Delta t^5D_{1j}^6 f_j^{(0)}=
-\frac{1}{\tau \Delta t} f_j^{(6)}, \label{Oeps6}
\end{eqnarray}
where
\begin{eqnarray}
\tau_2=&-&\tau+\frac{1}{2},\nonumber\\
\tau_3=&&\tau^2-\tau+\frac{1}{6},\nonumber\\
\tau_4=&-&\tau^3+\frac{3}{2}\tau^2-\frac{7}{12}\tau+\frac{1}{24},\nonumber\\
\tau_5=&&\tau^4-2\tau^3+\frac{5}{4}\tau^2-\frac{1}{4}\tau+\frac{1}{120},\nonumber\\
\tau_6=&-&\tau^5+\frac{5}{2}\tau^4-\frac{13}{6}\tau^3+\frac{3}{4}\tau^2-\frac{31}{360}\tau+\frac{1}{720}.\label{taus}
\end{eqnarray}

\subsection{Recovery of the Third-Order NPDE}

Since general LB models for the second-order NPDE have been
developed \cite{g,ShiGuo}, we only discuss how to recover NPDEs with
orders higher than two in this paper. Summing Eqs. (\ref{Oeps1}) and
(\ref{Oeps2}) over $j$, and using Eqs. (\ref{moments}) and
(\ref{Neq}), we obtain
\begin{eqnarray}
\sum_j D_{1j} f_j^{(0)} =\partial_{t_1} \phi +
\partial_{x_1}\Pi_1(\phi)= 0,\label{Oeps1E}\\
\partial_{t_2} \phi +\tau_2\Delta t
\sum_j D_{1j}^2 f_j^{(0)}= 0. \label{Oeps2E0}
\end{eqnarray}
Using Eqs. (\ref{moments}) and (\ref{Oeps1E}), and taking $\Pi_{20}$
as in Ref. \cite{ShiGuo} such that
$\partial_{t_1}\Pi_1+\partial_{x_1}\Pi_{20}=0$, that is
$\Pi_{20}=\int
\partial_{\phi} \Pi_1
\partial_{\phi} \Pi_1 d\phi$, we have
\begin{eqnarray}
\sum_j D_{1j}^2 f_j^{(0)} =\partial_{t_1}^2 \phi +
2\partial_{t_1}\partial_{x_1}\Pi_1+\partial_{x_1}^2(\Pi_{20}+\beta_2\Pi_2)
=\partial_{x_1}(\partial_{t_1}\Pi_1+\partial_{x_1}\Pi_{20})+\beta_2\partial_{x_1}^2\Pi_{2}=\beta_2\partial_{x_1}^2\Pi_{2},
\label{Q2}
\end{eqnarray}
then it follows from Eqs. (\ref{Oeps2E0}) and (\ref{Q2}) that
\begin{equation}
\partial_{t_2} \phi +\alpha_2\partial
_{x_1}^2 \Pi_2=0, \label{Oeps2E}
\end{equation} with
$\alpha_2=\Delta t \tau_2 \beta_2$.

Summing Eq. (\ref{Oeps3}) over $j$, and using Eqs. (\ref{moments}),
(\ref{Neq}), (\ref{Oeps1E}) and (\ref{Q2}), we obtain
\begin{eqnarray}
\partial_{t_3} \phi +\tau_3\Delta t^2 \sum_j D_{1j}^3 f_j^{(0)}= 0,
\label{Oeps3E0}
\end{eqnarray}
\begin{eqnarray}
\sum_j D_{1j}^3 f_j^{(0)} &=& \partial_{t_1}^3 \phi +
3\partial_{t_1}^2\partial_{x_1}\Pi_1+3\partial_{t_1}\partial_{x_1}^2(\Pi_{20}+\beta_2\Pi_2)+\partial_{x_1}^3(\Pi_{30}+\beta_3\Pi_3)
\nonumber\\&=&
2\partial_{t_1}^2\partial_{x_1}\Pi_1+3\partial_{t_1}\partial_{x_1}^2(\Pi_{20}+\beta_2\Pi_2)+\partial_{x_1}^3(\Pi_{30}+\beta_3\Pi_3)
 \nonumber\\&=&\partial_{x_1}^2(\partial_{t_1}(\Pi_{20}+3\beta_2\Pi_2)+\partial_{x_1}\Pi_{30})+\beta_3\partial_{x_1}^3\Pi_{3}.
\label{Q3}
\end{eqnarray}

If we take $\Pi_{30}$ such that
$\partial_{t_1}(\Pi_{20}+3\beta_2\Pi_2)+\partial_{x_1}\Pi_{30}=0$,
that is,
\begin{equation}
\Pi_{30}=\int\partial_\phi(\Pi_{20}+3\beta_2\Pi_2)\partial_\phi\Pi_1
d\phi=\int\left[(\partial_\phi
\Pi_1)^3+3\beta_2\partial_\phi\Pi_2\partial_\phi\Pi_1\right] d\phi,
\end{equation} then from Eqs. (\ref{Oeps3E0}) and (\ref{Q3}), we
have
\begin{equation}
\partial_{t_3}\phi+\alpha_3\partial_{x_1}^3\Pi_{3}=0,
\label{Oeps3E}
\end{equation} with $\alpha_3=\Delta
t^2\tau_3\beta_3$.

Combining Eqs. (\ref{Oeps1E}), (\ref{Oeps2E}) and (\ref{Oeps3E}),
the third-order NPDE is exactly recovered to order $O(\epsilon^3)$

\begin{equation}
\partial_t \phi+\partial_x \Pi_1 (\phi)+ \alpha_2 \partial_x^2 \Pi_2 (\phi)+ \alpha_3 \partial_x^3 \Pi_3 (\phi)=0,
\label{NPDE3}
\end{equation}
with $\alpha_2=\Delta t \tau_2\beta_2,\alpha_3=\Delta t^2
\tau_3\beta_3$.

\emph{Remark 1.} Note that there are no additional assumptions on
the present model for the third-order NPDEs with the form of Eq.
(\ref{NPDE}), and if we select $\Pi_{20}$ and $\Pi_{30}$ above the
third-order NPDEs are exactly recovered. The present model with a
D1Q4 or D1Q5 lattice can be used to simulate the third-order NPDE
(\ref{NPDE}) which contains some KdV-type equations.

\subsection{Recovery of the Fourth-Order NPDE}

Summing Eq. (\ref{Oeps4}) over $j$, and using Eqs. (\ref{moments}),
(\ref{Neq}), (\ref{Oeps1E}), (\ref{Q2}) and (\ref{Q3}), we obtain
\begin{eqnarray}
\partial_{t_4} \phi +3\tau_3\Delta t^2\partial_{t_2}\partial_{x_1}^2(\beta_2\Pi_2)+\tau_2\Delta t\partial_{t_2}^2 \phi+\tau_4\Delta t^3 \sum_j D_{1j}^4 f_j^{(0)}= 0.
\label{Oeps4E0}
\end{eqnarray}
\begin{eqnarray}
\sum_j D_{1j}^4 f_j^{(0)} &=& \partial_{t_1}^4 \phi
+4\partial_{t_1}^3\partial_{x_1}\Pi_1
+6\partial_{t_1}^2\partial_{x_1}^2(\Pi_{20}+\beta_2\Pi_2)+4\partial_{t_1}\partial_{x_1}^3(\Pi_{30}+\beta_3\Pi_3)+\partial_{x_1}^4(\Pi_{40}+\beta_4\Pi_4)
 \nonumber\\&=&3\partial_{t_1}^3\partial_{x_1}\Pi_1
+6\partial_{t_1}^2\partial_{x_1}^2(\Pi_{20}+\beta_2\Pi_2)+4\partial_{t_1}\partial_{x_1}^3(\Pi_{30}+\beta_3\Pi_3)+\partial_{x_1}^4(\Pi_{40}+\beta_4\Pi_4)
\nonumber\\&=&
\partial_{t_1}^2\partial_{x_1}^2(3\Pi_{20}+6\beta_2\Pi_2)+4\partial_{t_1}\partial_{x_1}^3(\Pi_{30}+\beta_3\Pi_3)+\partial_{x_1}^4(\Pi_{40}+\beta_4\Pi_4)
\nonumber\\&=&
\partial_{t_1}^2\partial_{x_1}^2\Pi_{20}+\partial_{t_1}\partial_{x_1}^3(2\Pi_{30}+4\beta_3\Pi_3)+\partial_{x_1}^4(\Pi_{40}+\beta_4\Pi_4)
\nonumber\\&=&
\partial_{x_1}^3(\partial_{t_1}(2\Pi_{30}-\tilde{\Pi}_{30}+4\beta_3\Pi_3)+\partial_{x_1}\Pi_{40})+\beta_4\partial_{x_1}^4\Pi_4,
\label{Q4}
\end{eqnarray}
where $\tilde{\Pi}_{30}$ satisfies that
$\partial_{t_1}\Pi_{20}+\partial_{x_1}\tilde{\Pi}_{30}=0$, that is,
$\tilde{\Pi}_{30}=\int\partial_\phi\Pi_{20}\partial_\phi\Pi_1
d\phi$.

If we take $\Pi_{40}$ such that
$\partial_{t_1}(2\Pi_{30}-\tilde{\Pi}_{30}+4\beta_3\Pi_3)+\partial_{x_1}\Pi_{40}=0$,
that is,
\begin{equation}
\Pi_{40}=\int\partial_\phi(2\Pi_{30}-\tilde{\Pi}_{30}+4\beta_3\Pi_3)\partial_\phi\Pi_1
d\phi=\int\left[(\partial_\phi
\Pi_1)^4+6\beta_2\partial_\phi\Pi_2(\partial_\phi\Pi_1)^2+4\beta_3\partial_\phi\Pi_3\partial_\phi\Pi_1\right]
d\phi,
\end{equation}
then from Eqs. (\ref{Oeps4E0}) and (\ref{Q4}) we have
\begin{equation}
\partial_{t_4}\phi+3\tau_3\beta_2\Delta t^2\partial_{t_2}\partial_{x_1}^2\Pi_2+\tau_2\Delta t\partial_{t_2}^2 \phi+\alpha_4\partial_{x_1}^4\Pi_{4}=0,
\label{Oeps4E1}
\end{equation} with $\alpha_4=\Delta
t^3\tau_4\beta_4$.

Note that there are two additional terms $3\tau_3\beta_2\Delta
t^2\partial_{t_2}\partial_{x_1}^2\Pi_2$ and $\tau_2\Delta
t\partial_{t_2}^2 \phi$ in Eq. (\ref{Oeps4E1}) should be removed
when correctly recovering Eq. (\ref{NPDE}). From Eq. (\ref{Oeps2E}),
we have
\begin{equation}
3\tau_3\beta_2\Delta
t^2\partial_{t_2}\partial_{x_1}^2\Pi_2+\tau_2\Delta
t\partial_{t_2}^2 \phi=(3\beta_2\tau_3\Delta
t^2-\alpha_2\tau_2\Delta
t)\partial_{t_2}\partial_{x_1}^2\Pi_2=(3\tau_3-\tau_2^2)\beta_2\Delta
t^2\partial_{t_2}\partial_{x_1}^2\Pi_2.
\end{equation}

If $\Pi_2=\phi$, then it follows from Eq. (\ref{Oeps2E}) that
\begin{equation}
3\tau_3\beta_2\Delta
t^2\partial_{t_2}\partial_{x_1}^2\Pi_2+\tau_2\Delta
t\partial_{t_2}^2 \phi=-\tau_2(3\tau_3-\tau_2^2)\beta_2^2\Delta
t^3\partial_{x_1}^4\phi.
\end{equation}

Combining above expression and Eq. (\ref{Q4}), we modify $\Pi_{40}$
as
\begin{equation}
\Pi_{40}=\int\left[(\partial_\phi
\Pi_1)^4+6\beta_2\partial_\phi\Pi_2(\partial_\phi\Pi_1)^2+4\beta_3\partial_\phi\Pi_3\partial_\phi\Pi_1
+A_{40}\right] d\phi, \label{Pi4}
\end{equation}
where $A_{40}=\frac{\tau_2(3\tau_3-\tau_2^2)\beta_2^2}{\tau_4}$, and
Eq. (\ref{Oeps4E1}) becomes

\begin{equation}
\partial_{t_4}\phi+\alpha_4\partial_{x_1}^4\Pi_{4}=0,
\label{Oeps4E}
\end{equation}
with $\alpha_4=\Delta t^3\tau_4\beta_4$.

Therefore, when $\Pi_2=\phi$, combining Eqs. (\ref{Oeps1E}),
(\ref{Oeps2E}), (\ref{Oeps3E}) and (\ref{Oeps4E}), the fourth-order
NPDE is exactly recovered to order $O(\epsilon^4)$
\begin{equation}
\partial_t \phi+\partial_x \Pi_1 (\phi)+ \alpha_2 \partial_x^2 \Pi_2 (\phi)+ \alpha_3 \partial_x^3 \Pi_3 (\phi)+ \alpha_4 \partial_x^4 \Pi_4 (\phi)=0,
\label{NPDE4}
\end{equation}
with $\alpha_2=\Delta t \tau_2\beta_2,\alpha_3=\Delta t^2
\tau_3\beta_3,\alpha_4=\Delta t^3 \tau_4\beta_4$.

\emph{Remark 2.} If $\alpha_2=0$ or $\Pi_2=0$, then we take
$\beta_2=0$ which leads to $A_{40}$=0. For this case, the
modification of $\Pi_{40}$ is not needed, and the fourth-order NPDE
is exactly recovered.

\emph{Remark 3.} The present model with a D1Q5 lattice can be used
to simulate the fourth-order NPDEs with the form as Eq. (\ref{NPDE})
which contains some Kuramoto-Sivashinsky-type equations. Recently,
following the idea in Ref. \cite{ChaiShiZheng} two similar LBGK
models with order $O(\epsilon^4)$ were given in Refs. \cite{MaCF1}
and \cite{MaCF2}, one for solving a class of three-order NPDEs which
were first solved by the LBGK model in Ref. \cite{ChaiShiZheng}, and
the other
 for the generalized Kuramoto-Sivashinsky
equation. It can be easily found that the models in Refs.
\cite{MaCF1} and \cite{MaCF2} do not satisfy the moments conditions
(\ref{moments}) for $m=4$. In fact, the so-called
\emph{amending-function} \cite{MaCF1, MaCF2} with order $O(\Delta
t^2)$ implies that there is a stronger assumption on the nonlinear
terms in EDF of the models. In addition, no comparisons were given
to show the \emph{higher-order} LBGK model in Ref. \cite{MaCF1}
superior to some lower-order LBGK models.

\subsection{Recovery of the Fifth-Order NPDE}

Summing Eq. (\ref{Oeps5}) over $j$, and using Eqs. (\ref{moments}),
(\ref{Neq}), (\ref{Oeps1E}), (\ref{Q2}), and (\ref{Q3}), we obtain
\begin{eqnarray}
\partial_{t_5} \phi +3\tau_3\Delta t^2\partial_{t_3}\partial_{x_1}^2(\beta_2\Pi_2)+2\tau_2\Delta t\partial_{t_2}\partial_{t_3} \phi
+4\tau_4\Delta t^3\partial_{t_2}\partial_{x_1}^3(\beta_3
\Pi_3)+\tau_5\Delta t^4 \sum_j D_{1j}^5 f_j^{(0)}= 0.
\label{Oeps5E0}
\end{eqnarray}

Using the similar procedure as Eq. (\ref{Q4}), we can obtain
\begin{eqnarray}
\sum_j D_{1j}^5 f_j^{(0)} = \beta_5\partial_{x_1}^5\Pi_5, \label{Q5}
\end{eqnarray}
with
\begin{equation}
\Pi_{50}=\int\left[(\partial_\phi\Pi_1)^5+10\beta_2\partial_\phi\Pi_2(\partial_\phi\Pi_1)^3+10\beta_3\partial_\phi\Pi_3(\partial_\phi\Pi_1)^2
+5\beta_4\partial_\phi\Pi_4\partial_\phi \Pi_1\right] d\phi,
\label{Pi50}
\end{equation}
and Eq. (\ref{Oeps5E0}) becomes
\begin{equation}
\partial_{t_5} \phi +3\tau_3\beta_2\Delta t^2\partial_{t_3}\partial_{x_1}^2\Pi_2
+2\tau_2\Delta t\partial_{t_2}\partial_{t_3}
\phi+4\tau_4\beta_3\Delta t^3\partial_{t_2}\partial_{x_1}^3
\Pi_3+\alpha_5\partial_{x_1}^5\Pi_5= 0, \label{Oeps5E1}
\end{equation}
with $\alpha_5=\Delta t^4\tau_5\beta_5$.

If $\Pi_2=\Pi_3=\phi$, we can modify $\Pi_{50}$ as
\begin{equation}
\Pi_{50}=\int\left[(\partial_\phi\Pi_1)^5+10\beta_2\partial_\phi\Pi_2(\partial_\phi\Pi_1)^3+10\beta_3\partial_\phi\Pi_3(\partial_\phi\Pi_1)^2+
5\beta_4\partial_\phi\Pi_4\partial_\phi \Pi_1+A_{50}\right] d\phi,
\label{Pi5}
\end{equation}
where
$A_{50}=\frac{(4\tau_2\tau_4+3\tau_3^2-2\tau_2^2\tau_3)\beta_2\beta_3}{\tau_5}$,
and Eq. (\ref{Oeps5E1}) becomes

\begin{equation}
\partial_{t_5} \phi +\alpha_5\partial_{x_1}^5\Pi_5= 0. \label{Oeps5E}
\end{equation}

Combining Eqs. (\ref{Oeps1E}), (\ref{Oeps2E}), (\ref{Oeps3E}),
(\ref{Oeps4E}) and (\ref{Oeps5E}), the fifth-order NPDE is exactly
recovered to order $O(\epsilon^5)$

\begin{equation}
\partial_t \phi+\partial_x \Pi_1 (\phi)+ \sum_{k=2}^5\alpha_k \partial_x^k \Pi_k (\phi)=0,
\label{NPDE5}
\end{equation}
with $\alpha_k=\Delta t^{k-1} \tau_k\beta_k, k=2,\ldots,5$.

\emph{Remark 4.} If $\alpha_2=0$ or $\alpha_3=0$, then $\beta_2=0$
or $\beta_3=0$ which leads to $A_{50}=0$. For this case, the
modification of $\Pi_{50}$ is not needed, and the fifth-order NPDE
is exactly recovered. The present model with a D1Q6 or D1Q7 lattice
can be used to simulate the fifth-order NPDE (\ref{NPDE}) ($m=5$)
which contains some Kawahara-like equations.

\subsection{Recovery of the Sixth-Order NPDE}

Summing Eq. (\ref{Oeps6}) over $j$, and using Eqs. (\ref{moments}),
(\ref{Neq}), (\ref{Oeps1E}), (\ref{Q2}), (\ref{Q3}), and (\ref{Q4}),
we obtain

\begin{eqnarray}
\partial_{t_6} \phi +3\tau_3\Delta t^2\partial_{t_4}\partial_{x_1}^2(\beta_2\Pi_2)+2\tau_2\Delta t\partial_{t_2}\partial_{t_4} \phi+
6\tau_4\Delta
t^3\partial_{t_2}^2\partial_{x_1}^2(\beta_2\Pi_2)+\tau_3\Delta
t^2\partial_{t_2}^3 \phi+\tau_2\Delta t\partial_{t_3}^2\phi
\nonumber\\+4\tau_4 \Delta t^3
\partial_{t_3}\partial_{x_1}^3(\beta_3\Pi_3)+5\tau_5 \Delta t^4
\partial_{t_2}\partial_{x_1}^4(\beta_4\Pi_4) +\tau_6\Delta t^5
\sum_j D_{1j}^6 f_j^{(0)}= 0. \label{Oeps6E0}
\end{eqnarray}

Using the similar procedure as Eq. (\ref{Q4}), we can obtain
\begin{eqnarray}
\sum_j D_{1j}^6 f_j^{(0)} = \beta_6\partial_{x_1}^6\Pi_6, \label{Q6}
\end{eqnarray}
with
\begin{equation}
\Pi_{60}=\int\left[(\partial_\phi\Pi_1)^6+15\beta_2\partial_\phi\Pi_2(\partial_\phi\Pi_1)^4+20\beta_3\partial_\phi\Pi_3(\partial_\phi\Pi_1)^3
+15\beta_4\partial_\phi\Pi_4(\partial_\phi
\Pi_1)^2+6\beta_5\partial_\phi\Pi_{5}\partial_\phi\Pi_1\right]
d\phi, \label{Pi60}
\end{equation}
and Eq. (\ref{Oeps6E0}) becomes
\begin{eqnarray}
\partial_{t_6} \phi +3\tau_3\beta_2\Delta t^2\partial_{t_4}\partial_{x_1}^2\Pi_2+2\tau_2\Delta t\partial_{t_2}\partial_{t_4} \phi+
6\tau_4\beta_2\Delta
t^3\partial_{t_2}^2\partial_{x_1}^2\Pi_2+\tau_3\Delta
t^2\partial_{t_2}^3 \phi+\tau_2\Delta t\partial_{t_3}^2\phi
\nonumber\\+4\tau_4 \beta_3\Delta t^3
\partial_{t_3}\partial_{x_1}^3\Pi_3+5\tau_5\beta_4 \Delta t^4
\partial_{t_2}\partial_{x_1}^4\Pi_4 +\alpha_6 \partial_{x_1}^6 \Pi_6= 0, \label{Oeps6E1}
\end{eqnarray}
with $\alpha_6=\Delta t^5\tau_6\beta_6$.

If $\Pi_2=\Pi_3=\Pi_4=\phi$, we can modify $\Pi_{60}$ as
\begin{equation}
\Pi_{60}=\int\left[(\partial_\phi\Pi_1)^6+15\beta_2\partial_\phi\Pi_2(\partial_\phi\Pi_1)^4+20\beta_3\partial_\phi\Pi_3(\partial_\phi\Pi_1)^3
+15\beta_4\partial_\phi\Pi_4(\partial_\phi
\Pi_1)^2+6\beta_5\partial_\phi\Pi_{5}\partial_\phi
\Pi_1+A_{60}\right] d\phi, \label{Pi6}
\end{equation}
where
$A_{60}=\frac{(5\tau_2\tau_5+3\tau_3\tau_4-2\tau_2^2\tau_4)\beta_2\beta_4+(4\tau_4-\tau_2\tau_3)\tau_3\beta_3^2
+(\tau_2\tau_3-6\tau_4)\tau_2^2\beta_2^3}{\tau_6}$, and Eq.
(\ref{Oeps6E1}) becomes

\begin{equation}
\partial_{t_6} \phi +\alpha_6\partial_{x_1}^6\Pi_6= 0. \label{Oeps6E}
\end{equation}

Combining Eqs. (\ref{Oeps1E}), (\ref{Oeps2E}), (\ref{Oeps3E}),
(\ref{Oeps4E}), (\ref{Oeps5E}), and  (\ref{Oeps6E}) the sixth-order
NPDE is exactly recovered to order $O(\epsilon^6)$

\begin{equation}
\partial_t \phi+\partial_x \Pi_1 (\phi)+ \sum_{k=2}^6\alpha_k \partial_x^k \Pi_k (\phi)=0,
\label{NPDE6}
\end{equation}
with $\alpha_k=\Delta t^{k-1} \tau_k\beta_k, k=2,\ldots,6$.

\emph{Remark 5.} If $\alpha_2=0$ and $\Pi_3=\phi$, then the
sixth-order NPDE is exactly recovered. The present model with a D1Q7
lattice can be used to simulate six-order NPDE (\ref{NPDE}) ($m=6$)
which also contains some Kawahara-like equations.

\section{Equilibrium Distribution Functions and Auxiliary Moments}

For a given NPDE of order $m$, $\Pi_{k0}$ can be obtained from the
related formula in above section ($2\leq k\leq m$), and the number
of discrete velocity directions is at least equal to $m+1$. So we
can use a D1Q5 LBGK model to solve the NPDE of order less than or
equal to 4, and a D1Q7 one to solve the NPDE of order less than or
equal to 6. A D1Q4 or D1Q6 one without the rest velocity can also be
used to the NPDE of order 3 or 5. In this section we give only the
EDFs and AMs for the LBGK models with D1Q5 and D1Q7 lattice,
respectively.

\subsection{Equilibrium Distribution Functions for LBGK Model with D1Q5 Lattice}

Denoting
$\bar{\Pi}_0=\phi,\bar{\Pi}_1=\Pi_1/c,\bar{\Pi}_k=(\Pi_{k0}+\beta_k\Pi_k)/c^k,k=2,3,4$,
the moments conditions (\ref{moments}) for $m=4$ are rewriten as
\begin{eqnarray}
 \sum_j e_j^k
f_j^{\mathrm{eq}}=\bar{\Pi}_k,k=0,\ldots,4 \label{moments5}
\end{eqnarray}
where $\{e_0,e_1,e_2,e_3,e_4\}=\{0,1,-1,2,-2\}$. Let

\begin{equation}
\vec{\mathbf{\Pi}}=[\bar{\Pi}_0,\bar{\Pi}_1,\ldots,\bar{\Pi}_4]^{T},
\vec{\mathbf{f}}^{\mathrm{eq}}=[f_0^{\mathrm{eq}},f_1^{\mathrm{eq}},\ldots,f_4^{\mathrm{eq}}]^{T}.
\end{equation}
From Eq. (\ref{moments5}), we have

\begin{equation}
\mathbf{M}_5 \vec{\mathbf{f}}^{\mathrm{eq}}=\vec{\Pi},
\end{equation}
where
\begin{displaymath}
\mathbf{M}_5=\left[
\begin{array} {rrrrr}
        1  &   1  &   1  &   1  &   1 \\
        0  &   1  &  -1  &   2  &  -2 \\
        0  &   1  &   1  &   4  &   4 \\
        0  &   1  &  -1  &   8  &  -8 \\
        0  &   1  &   1  &  16  &  16 \\
\end{array}
 \right],
\end{displaymath}

It is easy to find the inverse of $\mathbf{M}_5$

\begin{displaymath}
\mathbf{M}_5^{-1}=\frac{1}{24}\left[
\begin{array} {rrrrr}
        24 &   0  &  -30  &  0  &  6 \\
        0  &  16  &   16  & -4  & -4 \\
        0  & -16  &   16  &  4  & -4 \\
        0  &  -2  &   -1  &  2  &  1 \\
        0  &   2  &   -1  & -2  &  1 \\

\end{array}
 \right],
\end{displaymath}
thus
\begin{equation}
 \vec{\mathbf{f}}^{\mathrm{eq}}=\mathbf{M}_5^{-1}\vec{\Pi}.
\end{equation}
Therefore, the EDFs of the LBGK model with D1Q5 lattice can be
obtained as follows
\begin{eqnarray}
f_0^{\mathrm{eq}}&=&\left[4\phi-5\bar{\Pi}_2+\bar{\Pi}_4 \right]/4,\nonumber\\
f_1^{\mathrm{eq}}&=&\left[4(\bar{\Pi}_1+\bar{\Pi}_2)-\bar{\Pi}_3-\bar{\Pi}_4\right]/6,\nonumber\\
f_2^{\mathrm{eq}}&=&\left[4(-\bar{\Pi}_1+\bar{\Pi}_2)+\bar{\Pi}_3-\bar{\Pi}_4\right]/6,\nonumber\\
f_3^{\mathrm{eq}}&=&\left[-2(\bar{\Pi}_1-\bar{\Pi}_3)-\bar{\Pi}_2+\bar{\Pi}_4\right]/24,\nonumber\\
f_4^{\mathrm{eq}}&=&\left[2(\bar{\Pi}_1-\bar{\Pi}_3)-\bar{\Pi}_2+\bar{\Pi}_4\right]/24,\label{EDF5}.
\end{eqnarray}

\subsection{Equilibrium Distribution Functions for LBGK with D1Q7 Lattice}

Similarly, denoting
$\bar{\Pi}_0=\phi,\bar{\Pi}_1=\Pi_1/c,\bar{\Pi}_k=(\Pi_{k0}+\beta_k\Pi_k)/c^k,k=2,\ldots,6$,
the moments conditions (\ref{moments}) for $m=6$ are rewriten as
\begin{eqnarray}
 \sum_j e_j^k
f_j^{\mathrm{eq}}=\bar{\Pi}_k,k=0,\ldots,6 \label{moments7}
\end{eqnarray}
where $\{e_0,e_1,e_2,e_3,e_4,e_5,e_6\}=\{0,1,-1,2,-2,3,-3\}$. Let
\begin{equation}
\vec{\mathbf{\Pi}}=[\bar{\Pi}_0,\bar{\Pi}_1,\ldots,\bar{\Pi}_6]^{T},
\vec{\mathbf{f}}^{\mathrm{eq}}=[f_0^{\mathrm{eq}},f_1^{\mathrm{eq}},\ldots,f_6^{\mathrm{eq}}]^{T}.
\end{equation}
From Eq. (\ref{moments7}), we have
\begin{equation}
\mathbf{M}_7 \vec{\mathbf{f}}^{\mathrm{eq}}=\vec{\Pi},
\end{equation}
where
\begin{displaymath}
\mathbf{M}_7=\left[
\begin{array} {rrrrrrr}
        1  &   1  &   1  &   1  &   1 &   1  &  1 \\
        0  &   1  &  -1  &   2  &  -2 &   3  & -3 \\
        0  &   1  &   1  &   4  &   4 &   9  &  9  \\
        0  &   1  &  -1  &   8  &  -8 &  27  & -27 \\
        0  &   1  &   1  &  16  &  16 &  81  &  81 \\
        0  &   1  &  -1  &  32  & -32 &  243 & -243 \\
        0  &   1  &   1  &  64  &  64 &  729 &  729 \\
\end{array}
 \right],
\end{displaymath}

It is easy to find the inverse of $\mathbf{M}_7$

\begin{displaymath}
\mathbf{M}_7^{-1}=\frac{1}{720}\left[
\begin{array} {rrrrrrr}

              720 &    0 & -980 &    0 &  280 &   0 & -20 \\
                0 &  540 &  540 & -195 & -195 &  15 &  15 \\
                0 & -540 &  540 &  195 & -195 & -15 &  15 \\
                0 & -108 &  -54 &  120 &   60 & -12 &  -6 \\
                0 &  108 &  -54 & -120 &   60 &  12 &  -6 \\
                0 &   12 &    4 &  -15 &   -5 &   3 &   1 \\
                0 &  -12 &    4 &   15 &   -5 &  -3 &   1 \\
\end{array}
 \right],
\end{displaymath}
thus
\begin{equation}
 \vec{\mathbf{f}}^{\mathrm{eq}}=\mathbf{M}_7^{-1}\vec{\Pi}.
\end{equation}
Therefore, the EDFs of the LBGK model with D1Q7 lattice can be
obtained as follows
\begin{eqnarray}
f_0^{\mathrm{eq}}&=&\left[36\phi-49\bar{\Pi}_2+14\bar{\Pi}_4-\bar{\Pi}_6\right]/36,\nonumber\\
f_1^{\mathrm{eq}}&=&\left[36(\bar{\Pi}_1+\bar{\Pi}_2)-13(\bar{\Pi}_3+\bar{\Pi}_4)+\bar{\Pi}_5+\bar{\Pi}_6\right]/48,\nonumber\\
f_2^{\mathrm{eq}}&=&\left[36(-\bar{\Pi}_1+\bar{\Pi}_2)+13(\bar{\Pi}_3-\bar{\Pi}_4)-\bar{\Pi}_5+\bar{\Pi}_6\right]/48,\nonumber\\
f_3^{\mathrm{eq}}&=&\left[-18\bar{\Pi}_1-9\bar{\Pi}_2+20\bar{\Pi}_3+10\bar{\Pi}_4-2\bar{\Pi}_5-\bar{\Pi}_6\right]/120,\nonumber\\
f_4^{\mathrm{eq}}&=&\left[ 18\bar{\Pi}_1-9\bar{\Pi}_2-20\bar{\Pi}_3+10\bar{\Pi}_4+2\bar{\Pi}_5-\bar{\Pi}_6\right]/120,\nonumber\\
f_5^{\mathrm{eq}}&=&\left[ 12\bar{\Pi}_1+4\bar{\Pi}_2-15\bar{\Pi}_3- 5\bar{\Pi}_4+3\bar{\Pi}_5+\bar{\Pi}_6\right]/720,\nonumber\\
f_6^{\mathrm{eq}}&=&\left[-12\bar{\Pi}_1+4\bar{\Pi}_2+15\bar{\Pi}_3-
5\bar{\Pi}_4-3\bar{\Pi}_5+\bar{\Pi}_6\right]/720, \label{EDF7}
\end{eqnarray}

\subsection{Auxiliary Moments}

For a given NPDE of order $m$, $\Pi_k (1\leq k \leq m)$ are known,
and $\Pi_{k0} (2\leq k \leq m)$ can be obtained by using the formula
in Sec. II., then we can obtain the EDFs of related LBGK model with
suitable lattice, and the LBGK model is derived. For example, the
EDFs of D1Q5 or D1Q7 LBGK model can be computed by Eq. (\ref{EDF5}),
or Eq. (\ref{EDF7}). In simulations in the present work, two classes
of concrete NPDEs are considered, and the related AMs of these NPDEs
are given below.

(1) Sixth-order NPDEs

\begin{equation}
u_t + \alpha u u_x + \beta u^n u_x + \sum_{k=2}^6\alpha_k
\partial_x^k u=0, \label{Eq_order6}
\end{equation}
where $\alpha, \beta, n$ and $\alpha_k (2\leq k\leq 6)$ are
constants, and $n\geq 0$. Equation (\ref{Eq_order6}) contains many
well-known equations, such as the KdV-Burgers equation, (m)KdV
equation, Kuramoto-Sivashinsky equation, Kawahara equation, and some
of their extensions.

We can use a LBGK model with D1Q7 lattice to solve Eq.
(\ref{Eq_order6}), and the first seven moments are needed. Since
$\Pi_1=\frac{\alpha}{2}u^2+\frac{\beta}{n+1}u^{n+1},\Pi_k=u,k\geq
2$, we have $\Pi'_1=\alpha u+\beta u^n,\Pi'_k=1,k\geq 2$. Equation
(\ref{Eq_order6}) can be exactly recovered  by using the LBGK model
proposed above, and the exact expressions of AM functions $\Pi_{k0}
(2\leq k\leq 6)$ can be obtained as follows by using the formula in
Sec. II.

\begin{eqnarray}
\Pi_{20}&=&\int (\Pi'_1)^2 du,\nonumber\\
\Pi_{30}&=&\int \left[(\Pi'_1)^3+3\beta_2\Pi'_2\Pi'_1 \right]du=\int
(\Pi'_1)^3 du+3\beta_2 \Pi_1,\nonumber\\
\Pi_{40}&=&\int
\left[(\Pi'_1)^4+6\beta_2\Pi'_2(\Pi'_1)^2+4\beta_3\Pi'_3\Pi'_1+A_{40}
\right]du=\int(\Pi'_1)^4 du
+6\beta_2\Pi_{20}+4\beta_3\Pi_1+A_{40}u ,\nonumber\\
\Pi_{50}&=&\int
\left[(\Pi'_1)^5+10\beta_2\Pi'_2(\Pi'_1)^3+10\beta_3\Pi'_3(\Pi'_1)^2+5\beta_4\Pi'_4\Pi'_1+A_{50}
\right]du\nonumber\\&=&\int\left[(\Pi'_1)^5
+10\beta_2(\Pi'_1)^3 \right]du +10\beta_3\Pi_{20}+5\beta_4\Pi_1+A_{50}u ,\nonumber\\
\Pi_{60}&=&\int
\left[(\Pi'_1)^6+15\beta_2\Pi'_2(\Pi'_1)^4+20\beta_3\Pi'_3(\Pi'_1)^3+15\beta_4\Pi'_4(\Pi'_1)^2+6\beta_5\Pi'_5\Pi'_1+A_{60}
\right]du\nonumber\\&=&\int\left[(\Pi'_1)^6+15\beta_2(\Pi'_1)^4+20\beta_3(\Pi'_1)^3
\right]du +15\beta_4\Pi_{20}+6\beta_5\Pi_1+A_{60}u ,\label{Pi30t60}
\end{eqnarray}
where
\begin{eqnarray}
\alpha_k&=&\Delta t^{k-1}\tau_k \beta_k, 2\leq k\leq 6,\nonumber\\
A_{40}&=&\frac{\tau_2(3\tau_3-\tau_2^2)\beta_2^2}{\tau_4},\label{A40}\nonumber\\
A_{50}&=&\frac{(4\tau_2\tau_4+3\tau_3^2-2\tau_2^2\tau_3)\beta_2\beta_3}{\tau_5},\label{A50}\nonumber\\
A_{60}&=&\frac{(5\tau_2\tau_5+3\tau_3\tau_4-2\tau_2^2\tau_4)\beta_2\beta_4+(4\tau_4-\tau_2\tau_3)\tau_3\beta_3^2
+(\tau_2\tau_3-6\tau_4)\tau_2^2\beta_2^3}{\tau_6},\label{A60}
\end{eqnarray}
and it is easily obtained that
\begin{equation}
\int(\Pi'_1)^k du= \sum_{j=0}^{k} \int C_k^j (\alpha u)^{k-j}(\beta u^n)^{j} du
=u\left[\sum_{j=0}^{k}\frac{1}{jn+k+1-j} C_k^j (\alpha u)^{k-j}(\beta u^n)^{j}\right],2\leq k \leq 6.\label{intPi1k}\\
\end{equation}
For $k=2, 3, 4$ we have
\begin{eqnarray}
\int(\Pi'_1)^2 du &=& u\left[\frac{1}{3}(\alpha u)^2+\frac{2}{n+2}(\alpha u)(\beta u^n)+\frac{1}{2n+1}(\beta u^n)^2\right],\label{intPi12}\nonumber\\
\int(\Pi'_1)^3 du &=& u\left[\frac{1}{4}(\alpha u)^3+\frac{3}{n+3}(\alpha u)^2(\beta u^n)+\frac{3}{2n+2}(\alpha u)(\beta u^n)^2+\frac{1}{3n+1}(\beta u^n)^3\right],\label{intPi13}\nonumber\\
\int(\Pi'_1)^4 du &=& u\left[\frac{1}{5}(\alpha
u)^4+\frac{4}{n+4}(\alpha u)^3(\beta u^n)+\frac{6}{2n+3}(\alpha
u)^2(\beta u^n)^2 +\frac{4}{3n+2}(\alpha u)(\beta
u^n)^3+\frac{1}{4n+1}(\beta u^n)^4\right].\label{intPi14}
\end{eqnarray}

(2) Third-order NPDEs

\begin{equation}
u_t + \alpha u u_x + \beta u^n u_x + \alpha_2 (u^p)_{xx} + \alpha_3
(u^q)_{xxx}=0, \label{Eq_order3}
\end{equation}
where $\alpha, \beta$, $\alpha_2$, $\alpha_3$, $n$, $p$ and $q$  are
constants, and $n\geq 0$, $p\geq 1$, and $q\geq 1$. Eq.
(\ref{Eq_order3}) contains also many well-known equations, such as
the KdV-Burgers equation, (m)KdV equation, and K($p,q$) equation.

We can use a LBGK model with D1Q4 or D1Q5 lattice to solve Eq.
(\ref{Eq_order3}), and if the C-E expansion to order $O(\epsilon^3)$
is used, we need only to give the first four moments, while for the
C-E expansion to order $O(\epsilon^4)$, the first five moments are
needed. Since
$\Pi_1=\frac{\alpha}{2}u^2+\frac{\beta}{n+1}u^{n+1},\Pi_2=u^p$ and
$\Pi_3=u^q$, we can exactly recover Eq. (\ref{Eq_order3}) by using
the LBGK model proposed above, and the exact expressions of AM
functions $\Pi_{20}$, $\Pi_{30}$ and $\Pi_{40}$ can be obtained as
follows by using the formula in Sec. II.

\begin{eqnarray}
\Pi_{20}&=&\int (\Pi'_1)^2 du,\nonumber\\
\Pi_{30}&=&\int \left[(\Pi'_1)^3+3\beta_2\Pi'_2\Pi'_1 \right]du=\int
\left[(\Pi'_1)^3+3\beta_2 p u^{p-1}(\alpha u+\beta u^n)
\right]du\nonumber\\&=& \int
(\Pi'_1)^3 du+3\beta_2 p u^p\left[\frac{\alpha u}{p+1}+\frac{\beta u^n}{p+n}\right],\nonumber\\
\Pi_{40}&=&\int
\left[(\Pi'_1)^4+6\beta_2\Pi'_2(\Pi'_1)^2+4\beta_3\Pi'_3\Pi'_1+A_{40}
\right]du\nonumber\\&=&\int(\Pi'_1)^4 du +6\beta_2 p
u^p\left[\frac{(\alpha u)^2}{p+2}+\frac{2(\alpha u)(\beta
u^n)}{p+n+1}+\frac{(\beta u^n)^2}{p+2n}\right] +4\beta_3 q
u^q\left[\frac{\alpha u}{q+1}+\frac{\beta u^n}{q+n}\right]+
A_{40}u ,\label{3Pi20t40}
\end{eqnarray}
where $A_{40}$, $\int (\Pi'_1)^2 du$, $\int (\Pi'_1)^3 du$, and
$\int (\Pi'_1)^4 du$ can be computed by Eqs. (\ref{A40}) and
(\ref{intPi14}).

\section{Simulation Results}

To test the LBGK model proposed above, three classes of NPDEs with
exact solutions are simulated, including four
Kuramoto-Sivashinsky-like equations, three Kawahara-like equations
and two KdV-like equations. In all simulations, if not specified, we
use the nonequilibrium extrapolation scheme proposed by Guo {\it et
al.} \cite{gzs} to treat the exact boundary condition, and the
initial and boundary conditions of the test problems with analytical
solutions are determined by their analytical solutions. The D1Q5 and
D1Q7 LBGK models are used to simulate the test problems. The
following global relative error is used to measure the accuracy:
\begin{equation}
E=\frac{\sum_{j}
|\phi(\mathbf{x}_j,t)-\phi^*(\mathbf{x}_j,t)|}{\sum_{j}
|\phi^*(\mathbf{x}_j,t)|}\,,
\end{equation}
where $\phi$  and $\phi^*$  are the numerical solution and
analytical one, respectively, and the summation is taken over all
grid points.

The first four test problems are the fourth-order Kuramoto-Sivashinsky-type equations,
and three of them were simulated by the LBGK model in Ref. \cite{MaCF2}.
We use the D1Q5 LBGK model to simulate them and compare the proposed model with that in Ref. \cite{MaCF2}.

\emph{Example 4.1.} The Kuramoto-Sivashinsky equation \cite{XuShu}
\begin{equation}
u_t + uu_x + u_{xx} + u_{xxxx} =0\label{KSE1},
\end{equation}
with the exact solution
\begin{eqnarray}
u(x,t)= b+
\frac{15}{19}\sqrt{\frac{11}{19}}\left(-9\tanh(k(x-bt-x_0))+11\tanh^3(k(x-bt-x_0))\right),\label{SKSE1}
\end{eqnarray}
where $b, k, x_0$ are parameters.

In simulations, we set $b=5, k=\frac{1}{2}\sqrt{\frac{11}{19}}$, and
$x_0=-12$ as in Ref. \cite{MaCF2} for comparison. The simulation is
conducted in $[-30,30]$ with $\Delta x=0.1, \Delta t=0.01, 0.001$, and $0.0001$,
corresponding to $c=10, 100$, and 1000, respectively.
The errors are listed in Table I for different times,
where $\tau_{opt}$ is the optimal one corresponding to the minimal
error. We also present the regular shock profile wave propagation
for Eq. (\ref{KSE1}) with Eq. (\ref{SKSE1}) in Fig. 1. From the
table it can be seen that the errors of our model are smaller than
those of the model in Ref. \cite{MaCF2}, and the accuracy of our
model for $\Delta t=0.01$ is much better than that of the model in
Ref. \cite{MaCF2}. When $\Delta t$ is small enough, the effect of truncated errors of the model
in Ref. \cite{MaCF2} can be ignored, thus the difference between the present model
and that in Ref. \cite{MaCF2} is less.

\begin{table*}
\caption{Comparison of global relative errors at different times [
(1) $t=1$; (2) $t=2$; (3) $t=3$; (4) $t=4$ ].}
\begin{ruledtabular}
\begin{tabular}{ccccccccc}
  &   & & Present Model &  &  & & Model in Ref. \cite{MaCF2} & \\
&$(c,\tau_{opt})$ & $(10,5.99)$ & $(100,1.989)$ & $(1000,1.27)$ & & $(10,3.346)$ & $(100,1.998)$ & $(1000,1.2705)$ \\
\hline
 (1)&  &$9.6476\times 10^{-3}$&$2.9926\times 10^{-3}$&$6.0570\times 10^{-4}$& &$2.6571\times 10^{-2}$&$3.2655\times 10^{-3}$&$6.5581\times 10^{-4}$ \\
 (2)&  &$1.2962\times 10^{-2}$&$4.2992\times 10^{-3}$&$8.7427\times 10^{-4}$& &$3.7065\times 10^{-2}$&$5.3215\times 10^{-3}$&$1.1121\times 10^{-3}$ \\
 (3)&  &$1.7247\times 10^{-2}$&$5.5078\times 10^{-3}$&$1.1185\times 10^{-3}$& &$5.0615\times 10^{-2}$&$7.1611\times 10^{-3}$&$1.5426\times 10^{-3}$ \\
 (4)&  &$2.2122\times 10^{-2}$&$6.8650\times 10^{-3}$&$1.3738\times 10^{-3}$& &$8.0887\times 10^{-2}$&$8.9284\times 10^{-3}$&$1.9441\times 10^{-3}$ \\
\end{tabular}
\end{ruledtabular}
\end{table*}

\begin{figure*}
\centering
\includegraphics[width=6.5cm,height=4.5cm]{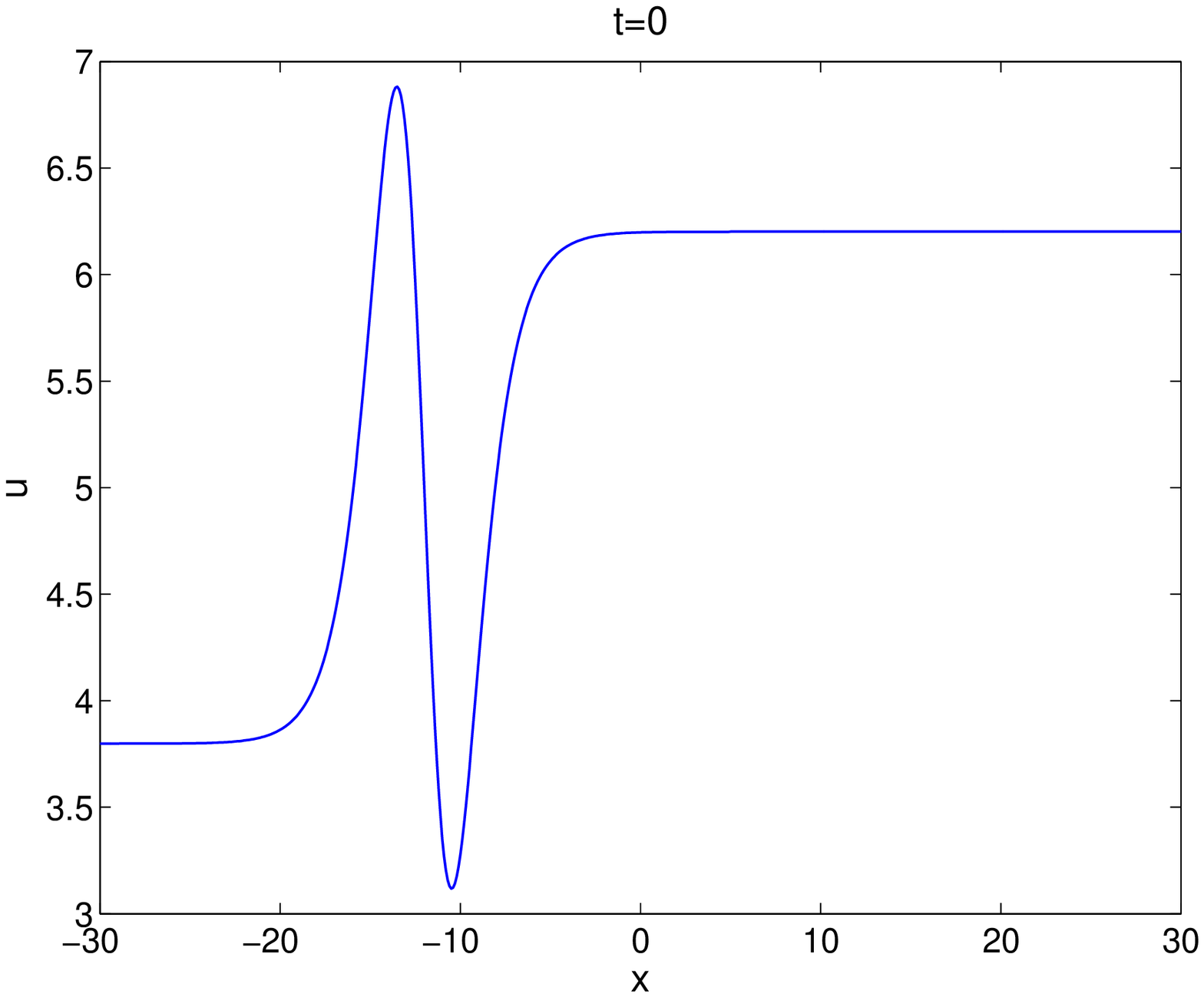}
\includegraphics[width=6.5cm,height=4.5cm]{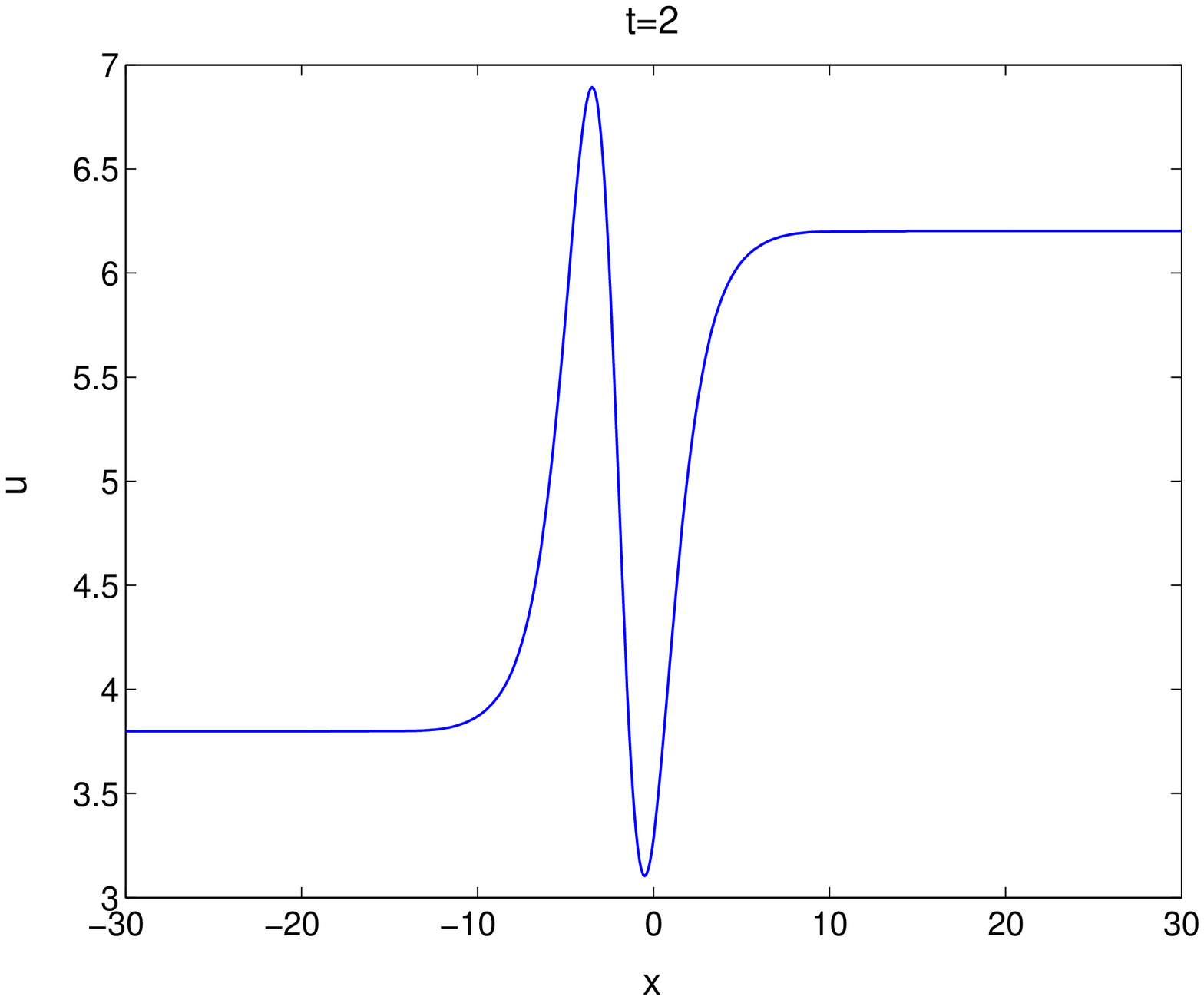}
\includegraphics[width=6.5cm,height=4.5cm]{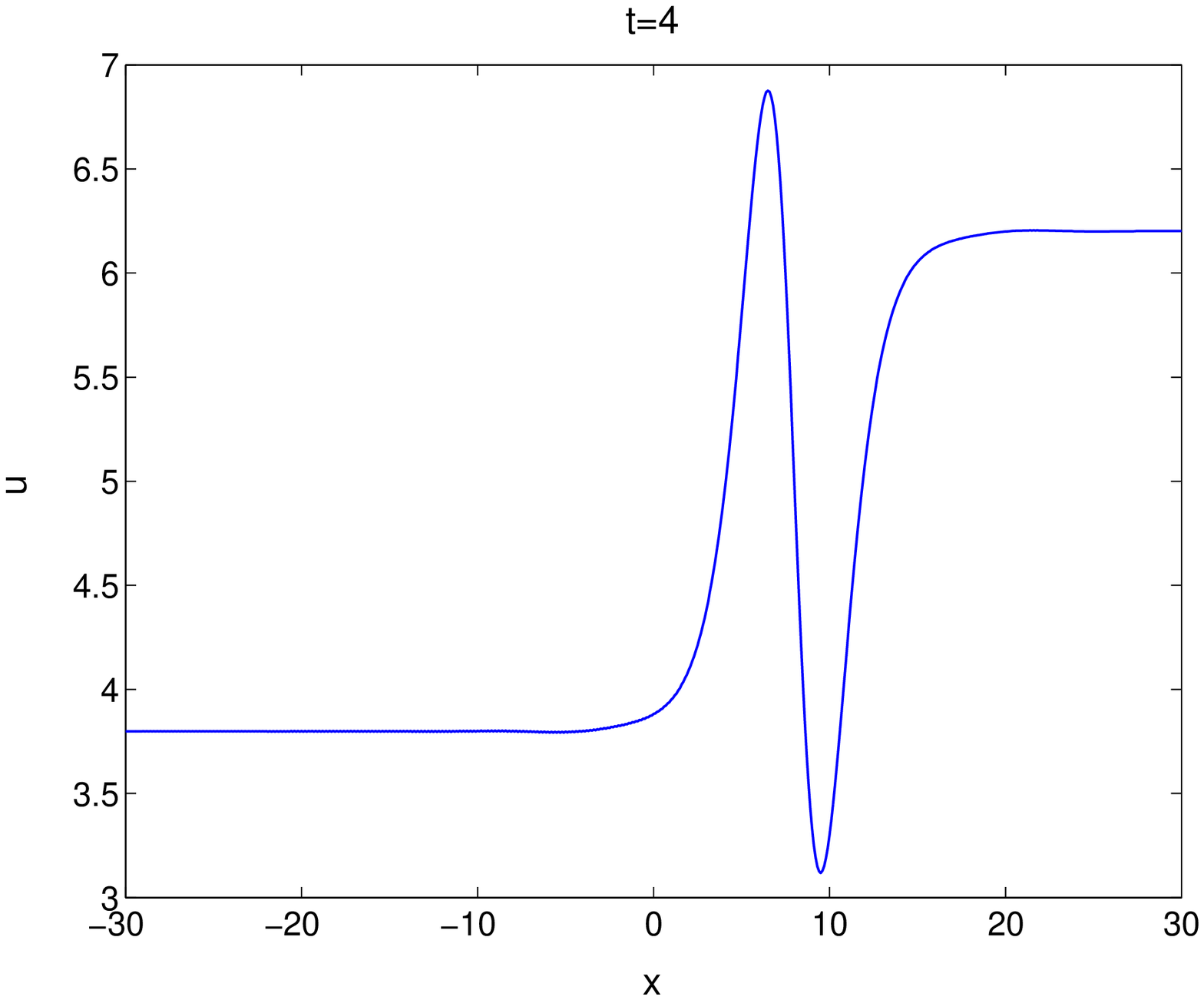}
\includegraphics[width=6.5cm,height=4.5cm]{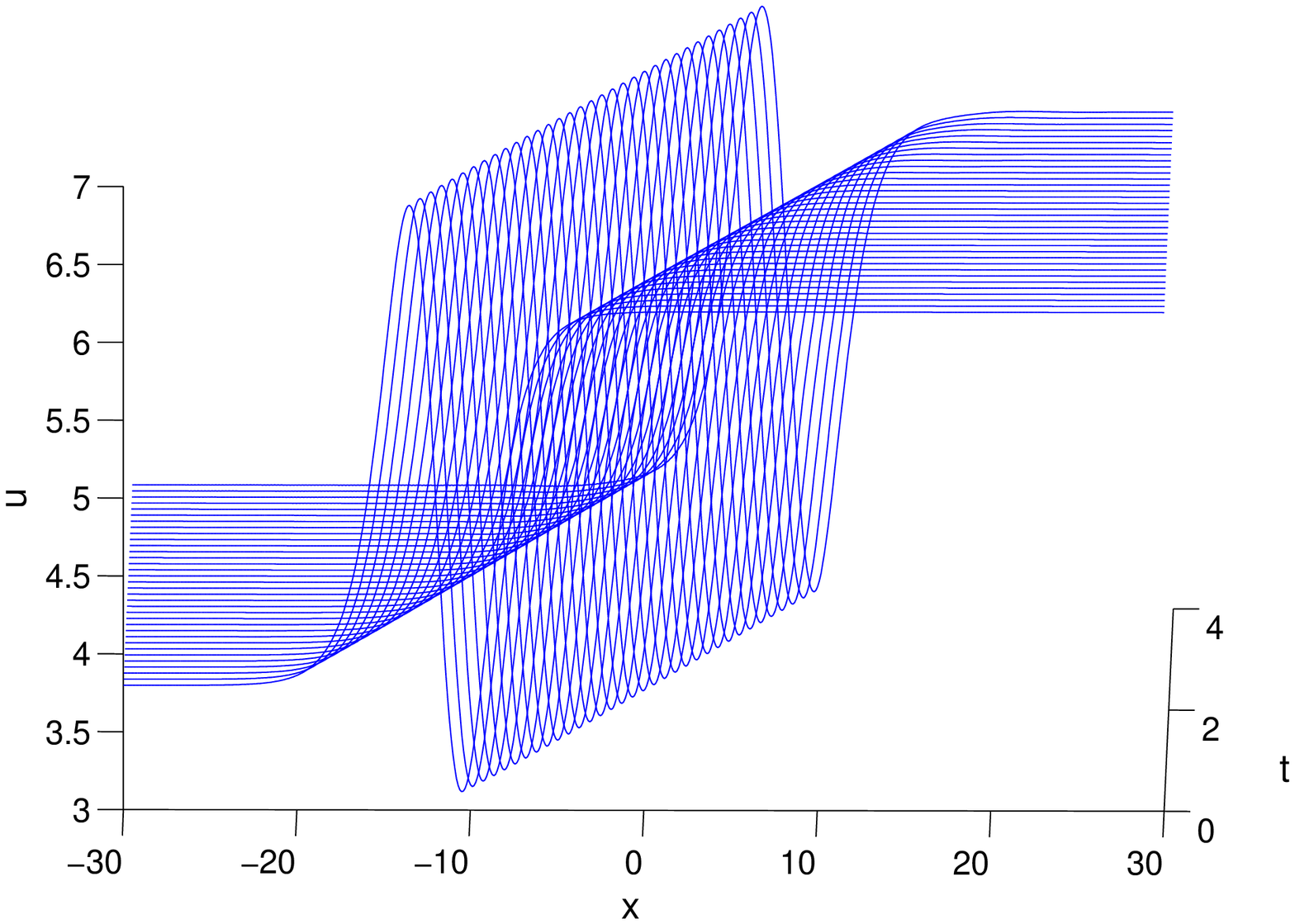}
\centerline{\hspace{0.2cm}} \caption{\label{fig:wide} (Color online)
The shock profile wave propagation of the KS equation (\ref{KSE1}),
$b = 5, k=\frac{1}{2}\sqrt{\frac{11}{19}}, x_0 = -12$. Exact
boundary condition in $[-30, 30]$, $\Delta x=0.1, \Delta t=0.0001$.}
\end{figure*}

\emph{Example 4.2.} The Kuramoto-Sivashinsky equation \cite{XuShu}
\begin{equation}
u_t + uu_x - u_{xx} + u_{xxxx} =0,\label{KSE2}
\end{equation}
with the exact solution
\begin{eqnarray}
u(x,t)= b+
\frac{15}{19\sqrt{19}}\left(-3\tanh(k(x-bt-x_0))+\tanh^3(k(x-bt-x_0))\right),
\end{eqnarray}
where $b, k, x_0$ are parameters.

In simulations, we set $b=5, k=\frac{1}{2\sqrt{19}}$, and $x_0=-25$
as in Ref. \cite{MaCF2} for comparison. The simulation is conducted in $[-50,50]$
with $\Delta x=0.1, \Delta t=0.01, 0.001$ and $0.0001$. The errors
are listed in Table II. for different times, and the regular shock profile wave
propagation for Eq. (\ref{KSE2}) is shown in Fig. 2. From the table it can be seen that the errors of our model are much smaller
than those of the model in Ref. \cite{MaCF2}.

\begin{table*}
\caption{Comparison of global relative errors at different times [
(1) $t=6$; (2) $t=8$; (3) $t=10$; (4) $t=12$ ].}
\begin{ruledtabular}
\begin{tabular}{ccccccccc}
  &   & & Present Model &  &  & & Model in Ref. \cite{MaCF2} & \\
&$(c,\tau_{opt})$ & $(10,4.569)$ & $(100,2.076)$ & $(1000,1.277)$ & & $(10,4.0)$ & $(100,2.063)$ & $(1000,1.277)$ \\
\hline
 (1)&  &$2.8486\times 10^{-5}$&$1.5343\times 10^{-6}$&$2.7448\times 10^{-7}$&  &$1.2313\times 10^{-3}$&$5.6085\times 10^{-5}$&$7.6750\times 10^{-6}$ \\
 (2)&  &$3.1775\times 10^{-5}$&$1.6534\times 10^{-6}$&$2.9535\times 10^{-7}$&  &$1.5818\times 10^{-3}$&$6.9643\times 10^{-5}$&$9.3058\times 10^{-6}$ \\
 (3)&  &$3.3937\times 10^{-5}$&$1.7189\times 10^{-6}$&$3.0741\times 10^{-7}$&  &$1.9018\times 10^{-3}$&$8.1373\times 10^{-5}$&$1.0640\times 10^{-5}$ \\
 (4)&  &$3.4934\times 10^{-5}$&$1.8264\times 10^{-6}$&$4.2625\times 10^{-7}$&  &$2.1886\times 10^{-3}$&$9.1494\times 10^{-5}$&$1.1661\times 10^{-5}$ \\
\end{tabular}
\end{ruledtabular}
\end{table*}

\begin{figure*}
\centering
\includegraphics[width=6.5cm,height=4.5cm]{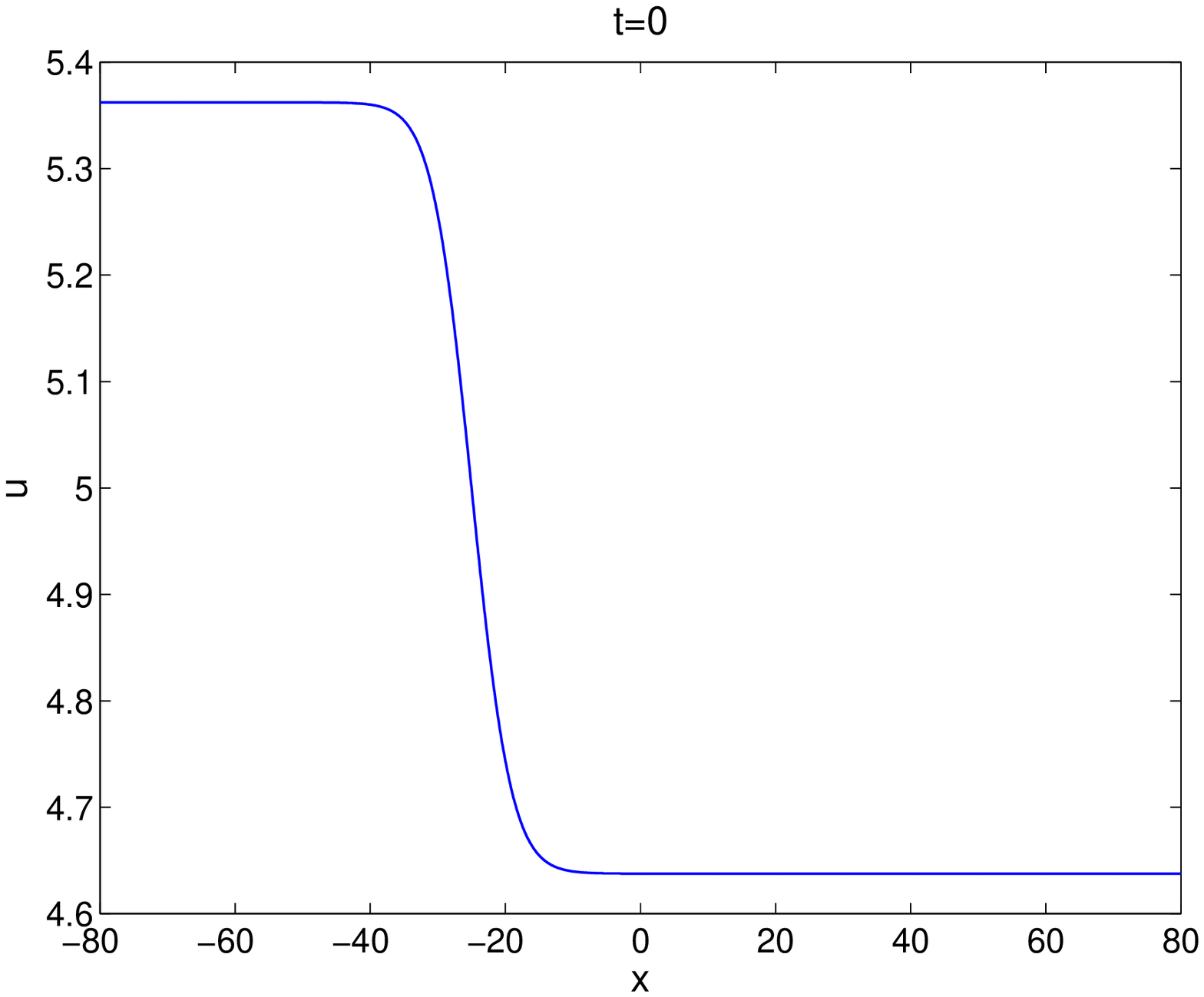}
\includegraphics[width=6.5cm,height=4.5cm]{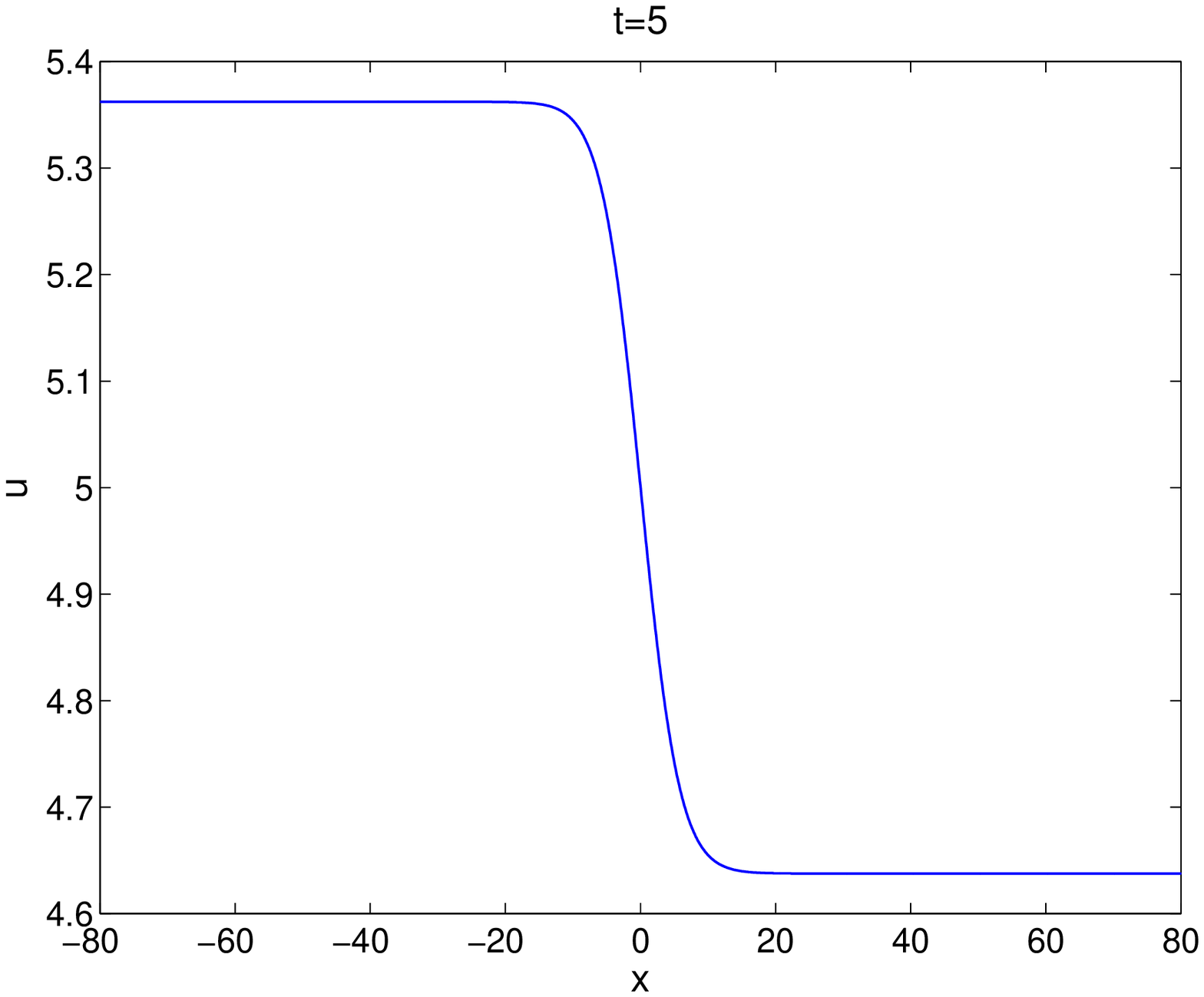}
\includegraphics[width=6.5cm,height=4.5cm]{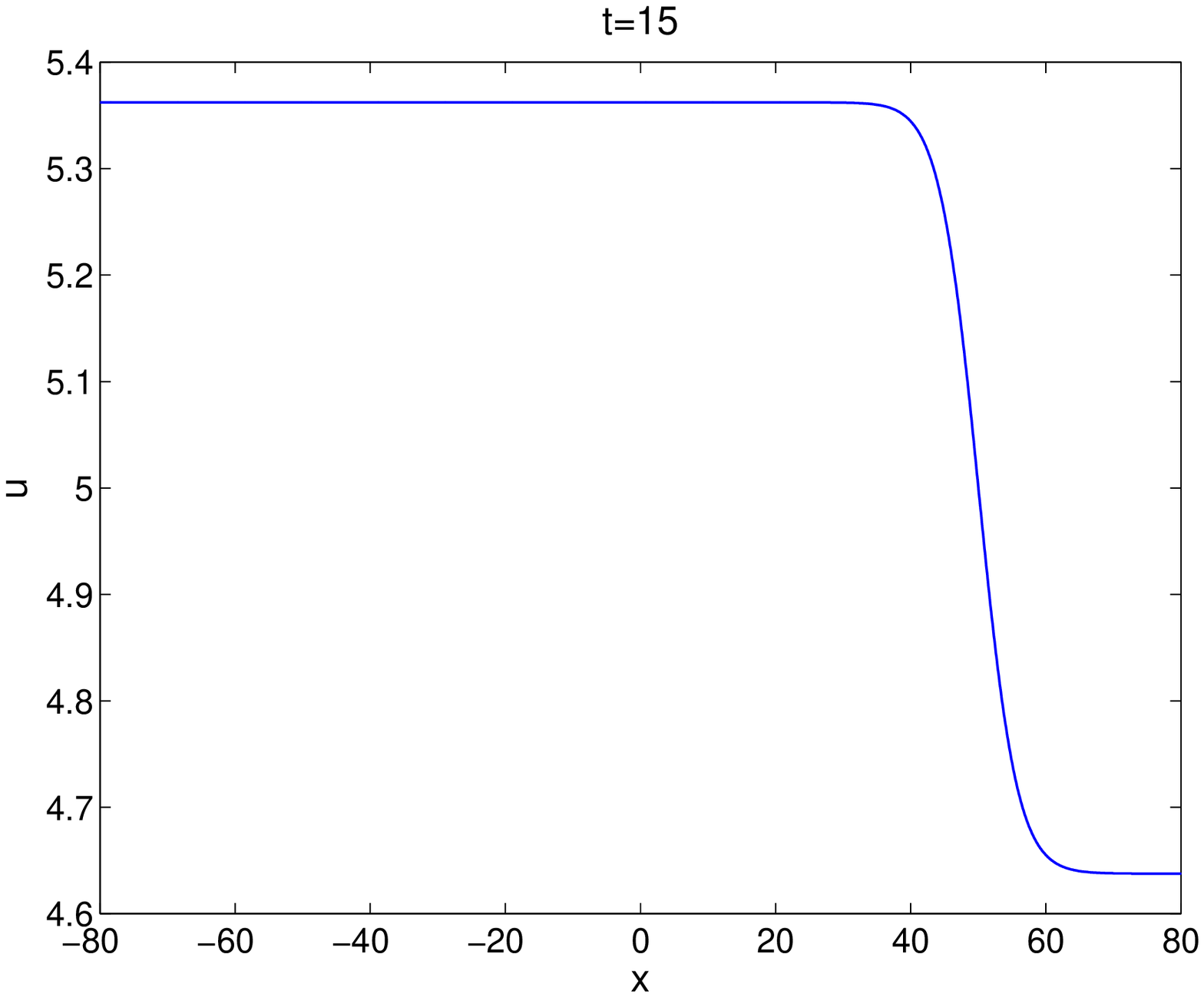}
\includegraphics[width=6.5cm,height=4.5cm]{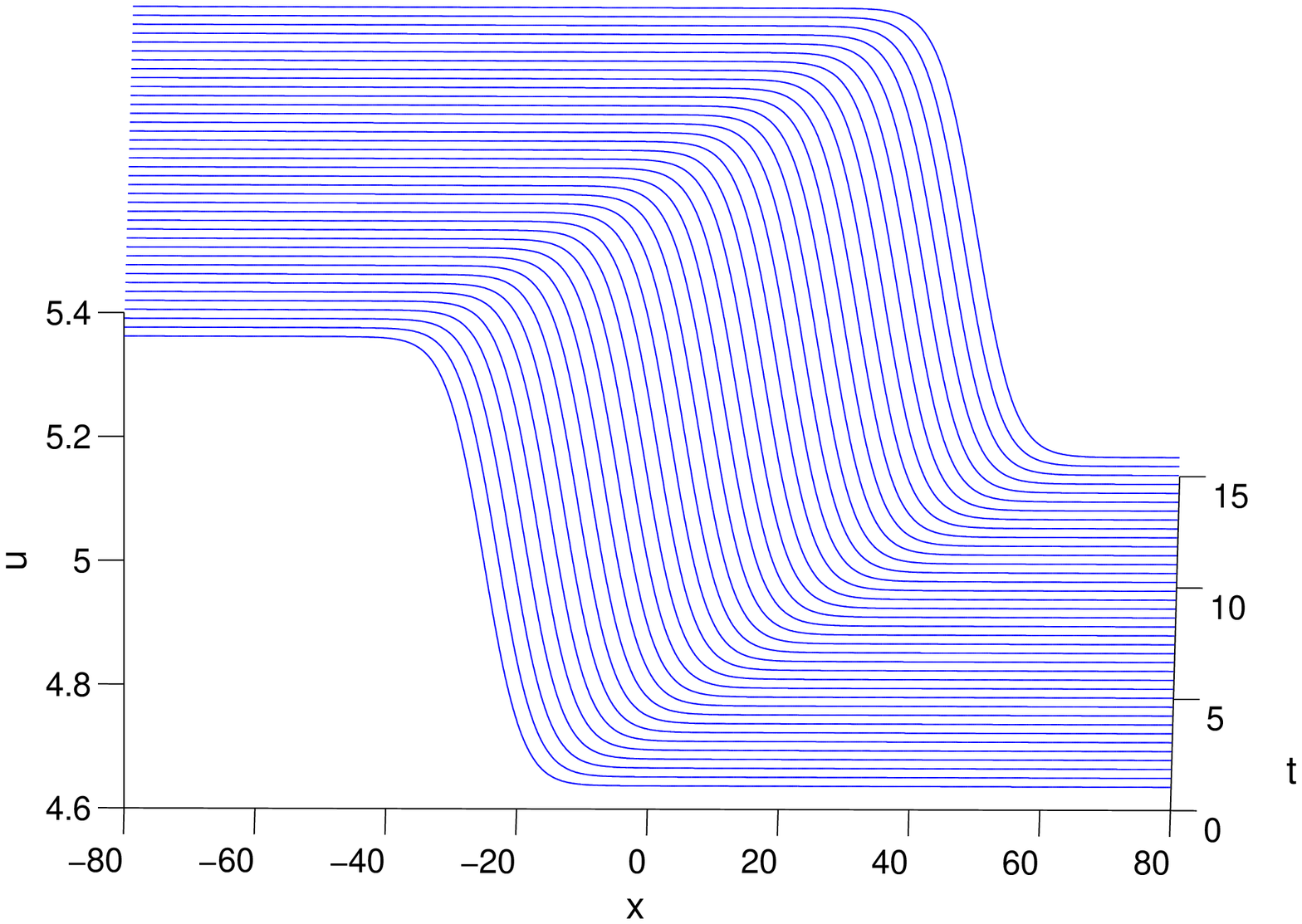}
\centerline{\hspace{0.2cm}} \caption{\label{fig:wide} (Color online)
The shock profile wave propagation of the KS equation (\ref{KSE2}),
$b = 5, k=\frac{1}{2\sqrt{19}}, x_0 = -25$. Exact boundary condition
in $[-80, 80]$, $\Delta x=0.1, \Delta t=0.01$.}
\end{figure*}

\emph{Example 4.3.} The generalized Kuramoto-Sivashinsky equation
\cite{XuShu}
\begin{equation}
u_t + uu_x + u_{xx} +\sigma u_{xxx} + u_{xxxx} =0,\label{GKSE1}
\end{equation}
with the exact solution
\begin{eqnarray}
u(x,t)=
b+9-15\left(\tanh(k(x-bt-x_0))+\tanh^2(k(x-bt-x_0))-\tanh^3(k(x-bt-x_0))\right),
\end{eqnarray}
where $\sigma, b, k, x_0$ are parameters.

In simulations, we set $\sigma=4, b=6, k=\frac{1}{2}$, and $x_0=-10$
as in Ref. \cite{MaCF2} for comparison. The simulation is performed
in $[-30,30]$ with $\Delta x=0.1, \Delta t=0.01, 0.001$ and
$0.0001$. Table III gives the errors of numerical solution at
different times. We also present the solitary wave propagation for
Eq. (\ref{GKSE1}) is shown in Fig. 3. From the table it can be seen
that the errors of our model for the smallest $\Delta t$ are larger than
those of the model in Ref. \cite{MaCF2}, while the accuracy and
stability of the present model for larger $\Delta t$ are better than
those of the model in Ref. \cite{MaCF2}. It is noted that the
accuracy of both models for Eq. (\ref{GKSE1}) is much larger than
for Eqs. (\ref{KSE1}) and (\ref{KSE2}).

\begin{table*}
\caption{Comparison of global relative errors at different times [
(1) $t=1$; (2) $t=2$; (3) $t=3$; (4) $t=4$; '-' means that the
scheme is divergent ].}
\begin{ruledtabular}
\begin{tabular}{ccccccccc}
  &   & & Present Model &  &  & & Model in Ref. \cite{MaCF2} & \\
&$(c,\tau_{opt})$ & $(10,7.082)$ & $(100,9.89)$ & $(1000,1.267)$ & & $(10,3.47)$ & $(100,1.975)$ & $(1000,1.267486)$ \\
\hline
 (1)&  &$4.1701\times 10^{-1}$&$5.2017\times 10^{-2}$&$5.1020\times 10^{-2}$&  &$9.7859\times 10^{-1}$&$1.3802\times 10^{-1}$&$2.6054\times 10^{-2}$ \\
 (2)&  &$1.2376\times 10^{-0}$&$6.9440\times 10^{-2}$&$5.6700\times 10^{-2}$&  &  -                   &$1.4077\times 10^{-1}$&$2.8329\times 10^{-2}$ \\
 (3)&  &$2.5757\times 10^{-0}$&$9.7967\times 10^{-2}$&$5.1337\times 10^{-2}$&  &  -                   &$1.7050\times 10^{-1}$&$2.6802\times 10^{-2}$ \\
 (4)&  &$3.4682\times 10^{-0}$&$1.6776\times 10^{-1}$&$6.5639\times 10^{-2}$&  &  -                   &$3.1488\times 10^{-1}$&$3.5225\times 10^{-2}$ \\
\end{tabular}
\end{ruledtabular}
\end{table*}

\begin{figure*}
\centering
\includegraphics[width=6.5cm,height=4.5cm]{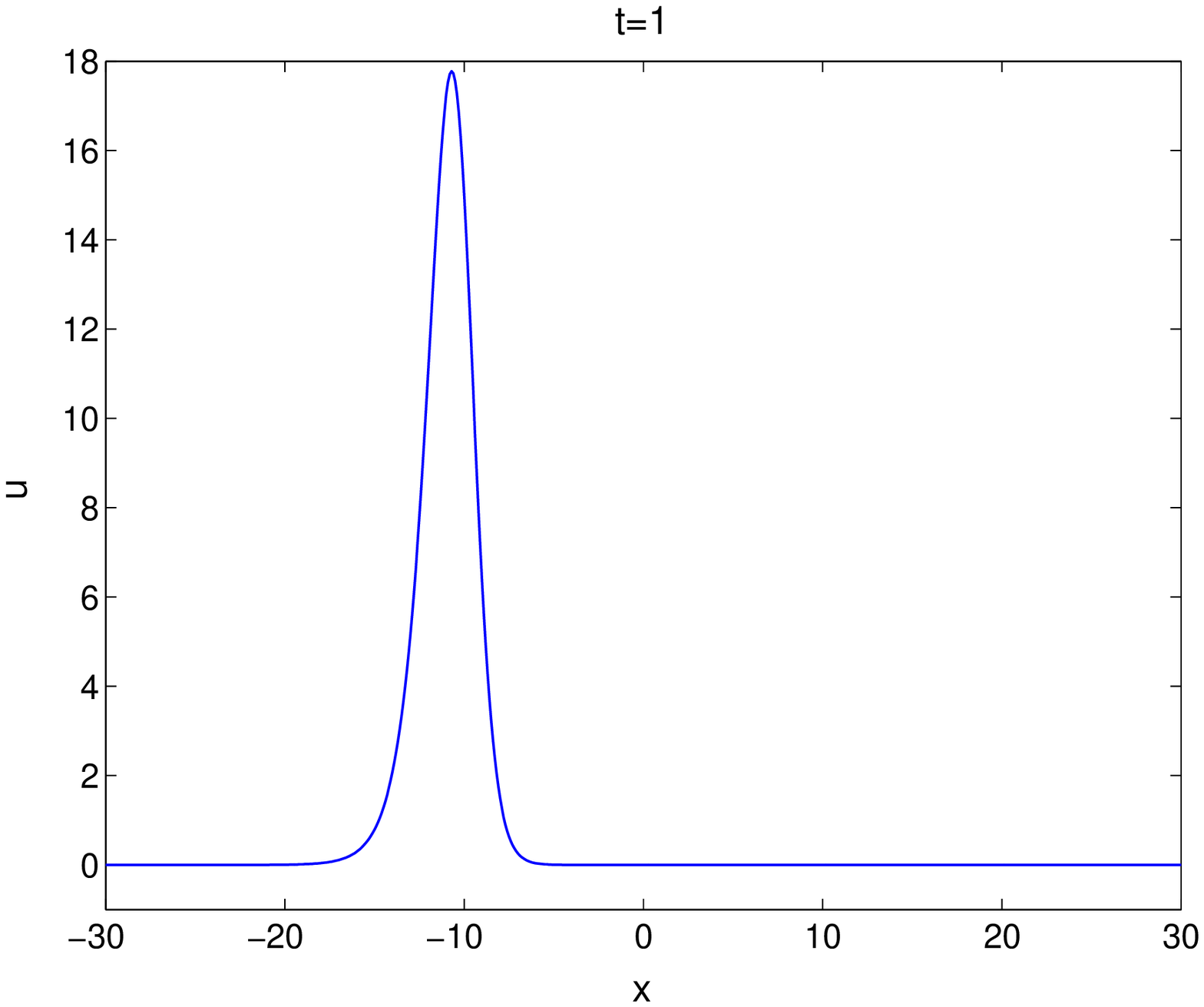}
\includegraphics[width=6.5cm,height=4.5cm]{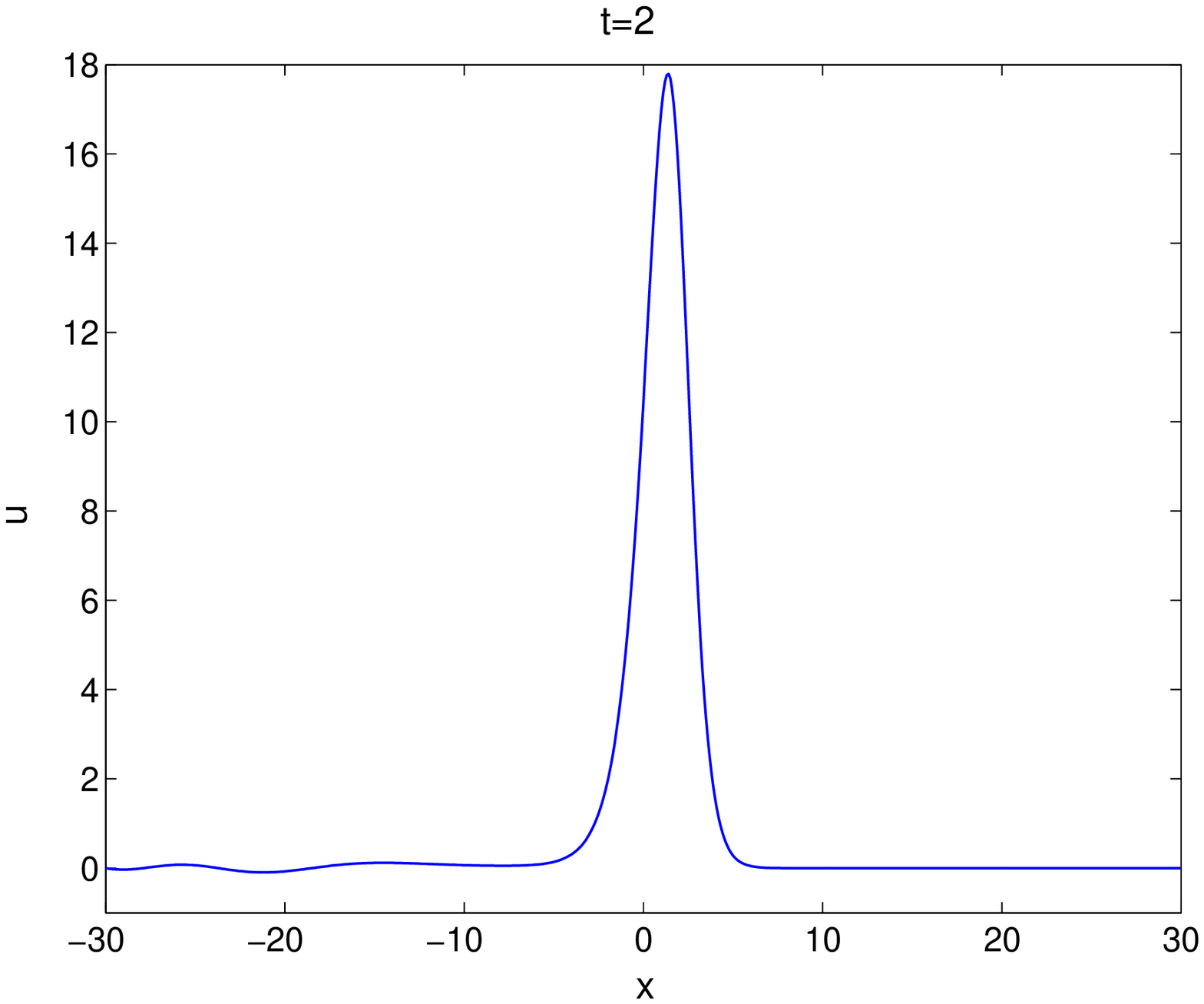}
\includegraphics[width=6.5cm,height=4.5cm]{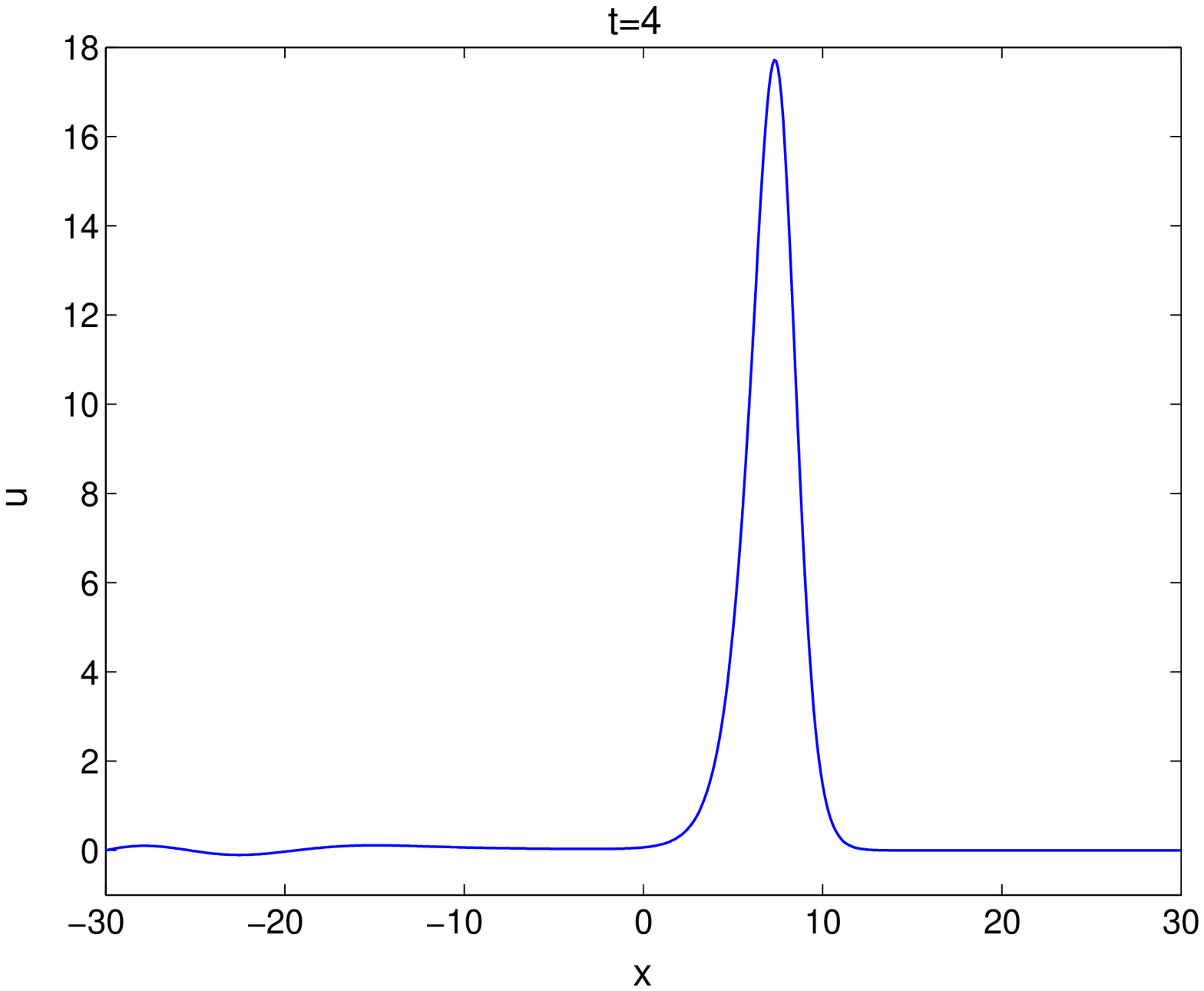}
\includegraphics[width=6.5cm,height=4.5cm]{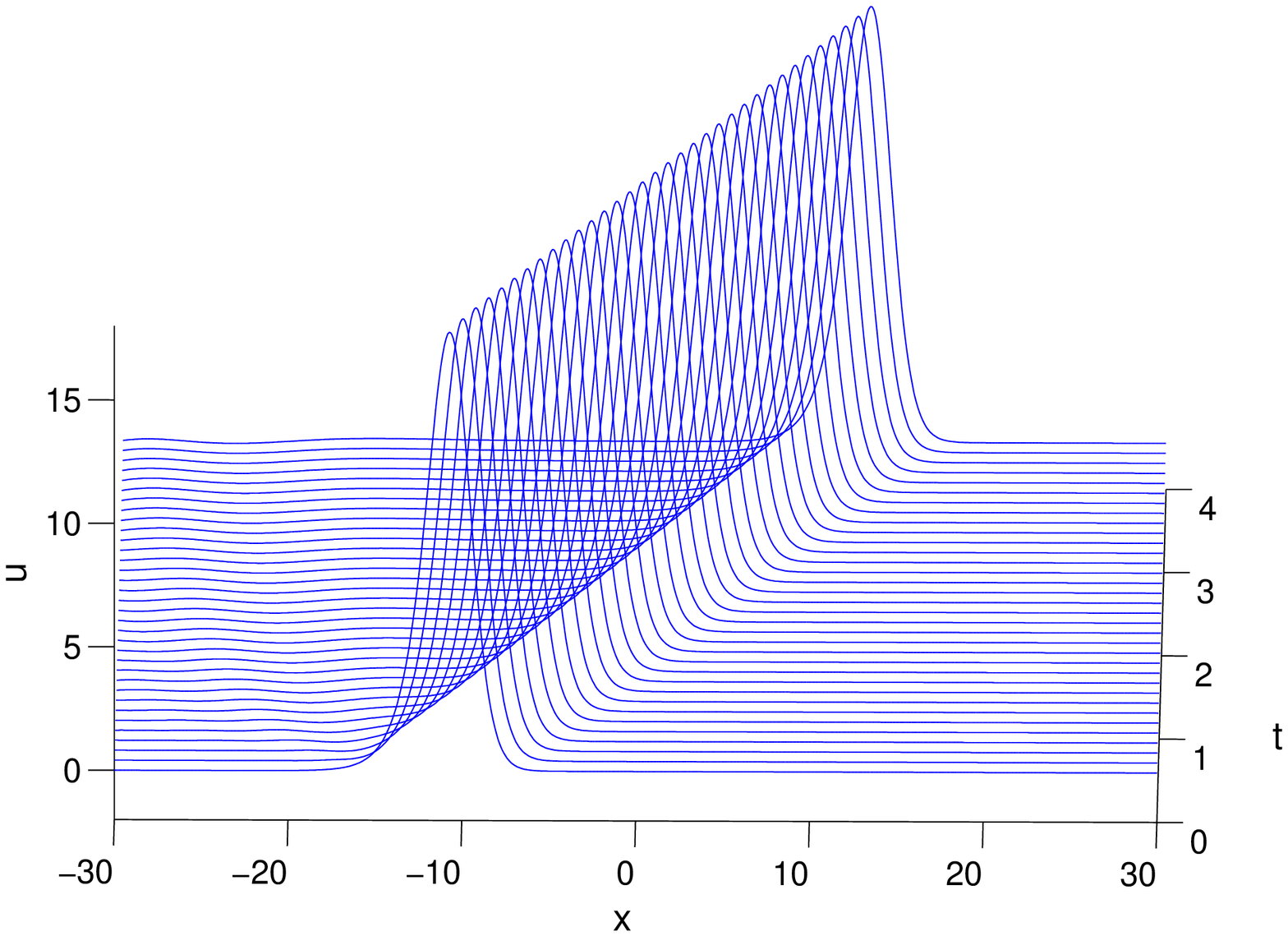}
\centerline{\hspace{0.2cm}} \caption{\label{fig:wide} (Color online)
The solitary wave propagation of the GKS equation (\ref{GKSE1}), $b
= 6, \sigma =4, k=\frac{1}{2}, x_0 = -10$. Exact boundary condition
in $[-30, 30]$, $\Delta x=0.1, \Delta t=0.0001$.}
\end{figure*}

\emph{Example 4.4.} The generalized Kuramoto-Sivashinsky equation
\cite{zhangzhang}
\begin{equation}
u_t + 3u^3u_x + a u_{xx} -b u_{xxx} +  u_{xxxx} =0,\label{GKSE2}
\end{equation}
with the exact solution
\begin{eqnarray}
u(x,t)=
\frac{\sqrt{3}b}{2\sqrt{2}}\tanh\left[\frac{\sqrt{3}b}{4\sqrt{2}}\left(x-x_0-\frac{29b^3}{144}t\right)+\frac{C}{2}\right]+\frac{b}{6},
\end{eqnarray}
where $a, b, C$ are constants.

In simulations, we set $a=1, b=1$, and $C=1$. The simulation is
performed in $[-30,30]$ with $x_0=0, \Delta x=0.1, \Delta t=0.01,
0.001$ and $0.0001$, and both D1Q5 and D1Q7 LBGK models are used.
Table IV gives the errors of numerical solution at different times,
and the regular shock profile wave propagation for Eq. (\ref{GKSE2})
is shown in Fig. 4. From the table It found that the numerical
solutions are agree well with the analytic ones, and the accuracy of
D1q7 model is much better than that of D1Q5 one for smaller $\Delta
t$.

\begin{table*}
\caption{Comparison of global relative errors at different times [
(1) $t=1$; (2) $t=2$; (3) $t=3$; (4) $t=4$ ].}
\begin{ruledtabular}
\begin{tabular}{ccccccccc}
  &   & & D1Q5 Model &  &  & & D1Q7 Model & \\
&$(c,\tau_{opt})$ & $(10,3.32)$ & $(100,2.0)$ & $(1000,1.27)$ & & $(10,4.14)$ & $(100,2.31)$ & $(1000,1.40)$ \\
\hline
 (1)&  &$1.4921\times 10^{-3}$&$3.7819\times 10^{-4}$&$7.5629\times 10^{-5}$&  &$1.3234\times 10^{-3}$&$8.7735\times 10^{-5}$&$4.2032\times 10^{-6}$ \\
 (2)&  &$3.1612\times 10^{-3}$&$7.8311\times 10^{-4}$&$1.5509\times 10^{-4}$&  &$2.6053\times 10^{-3}$&$1.5272\times 10^{-4}$&$6.9690\times 10^{-6}$ \\
 (3)&  &$5.0988\times 10^{-3}$&$1.2215\times 10^{-3}$&$2.4006\times 10^{-4}$&  &$4.1570\times 10^{-3}$&$2.2473\times 10^{-4}$&$1.0021\times 10^{-5}$ \\
 (4)&  &$7.2939\times 10^{-3}$&$1.6930\times 10^{-3}$&$3.3080\times 10^{-4}$&  &$6.0013\times 10^{-3}$&$3.0868\times 10^{-4}$&$1.3604\times 10^{-5}$ \\
\end{tabular}
\end{ruledtabular}
\end{table*}

\begin{figure*}
\centering
\includegraphics[width=6.5cm,height=4.5cm]{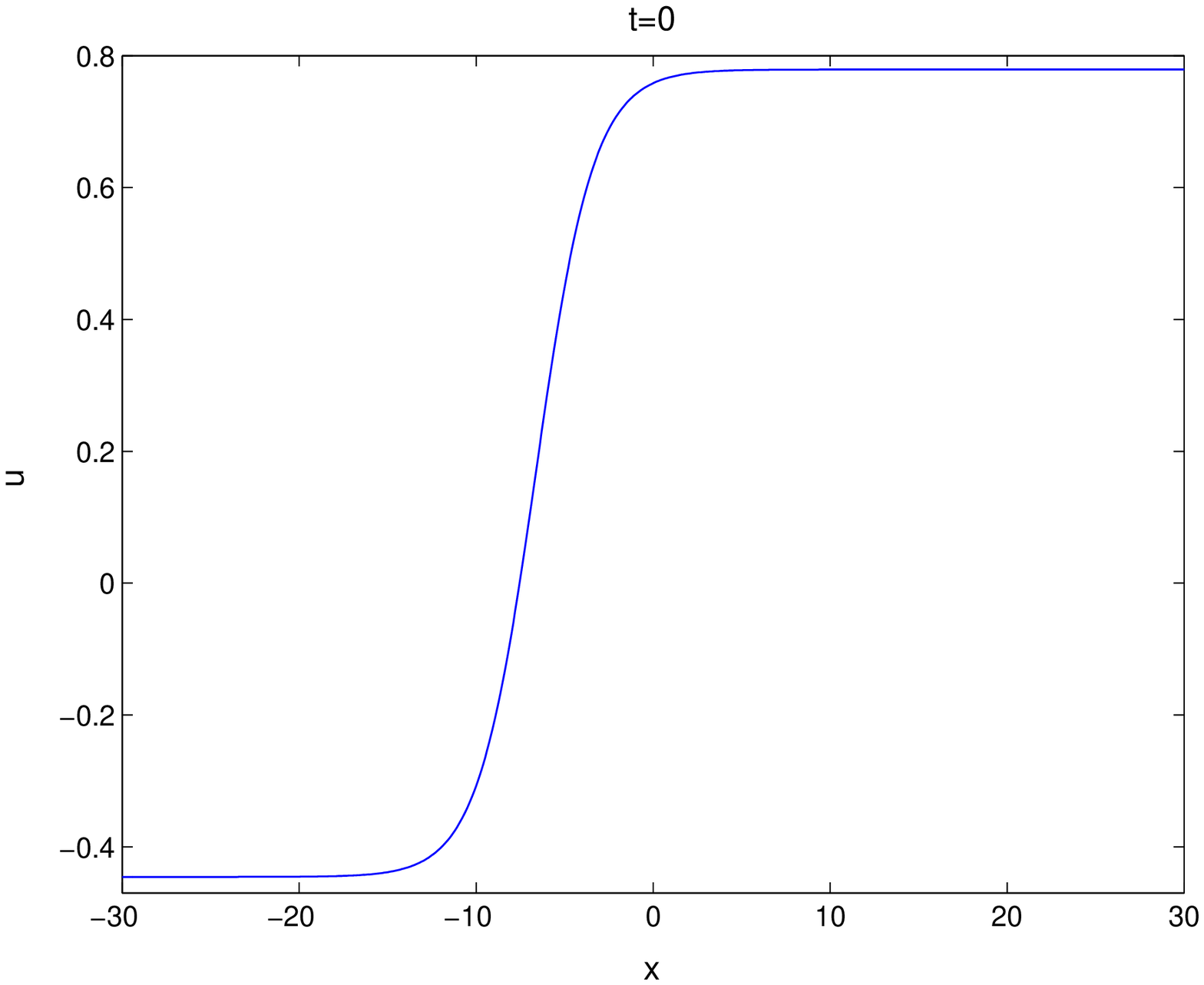}
\includegraphics[width=6.5cm,height=4.5cm]{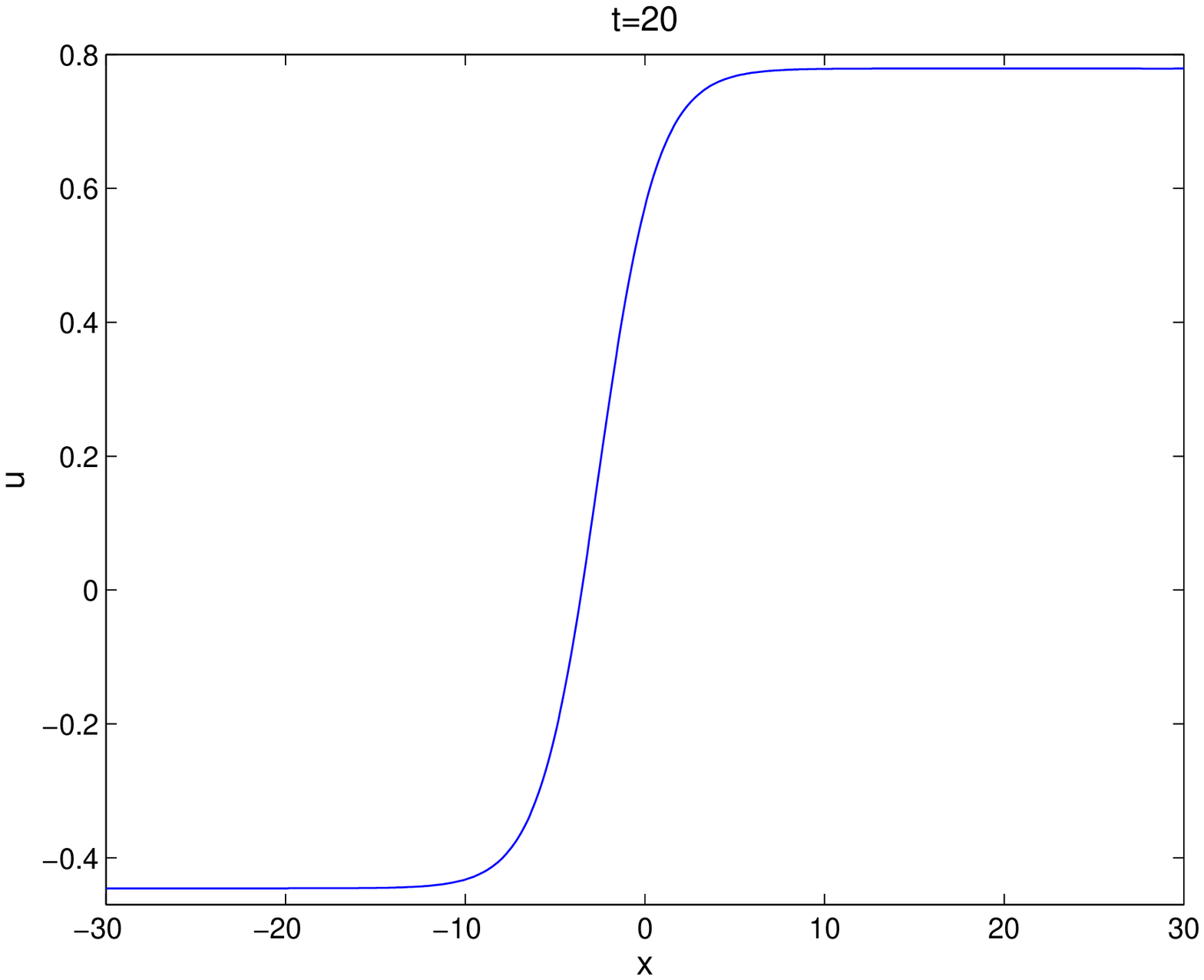}
\includegraphics[width=6.5cm,height=4.5cm]{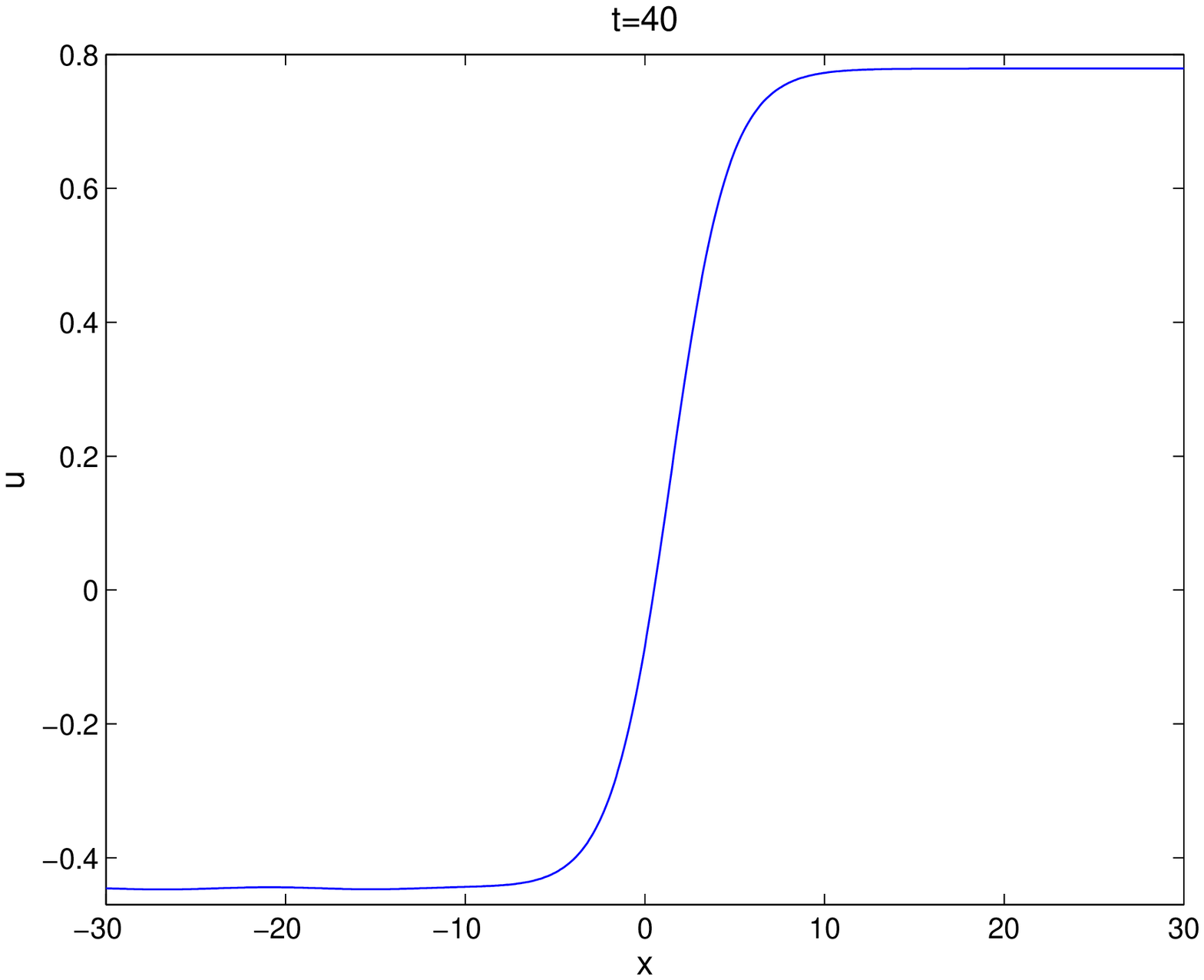}
\includegraphics[width=6.5cm,height=4.5cm]{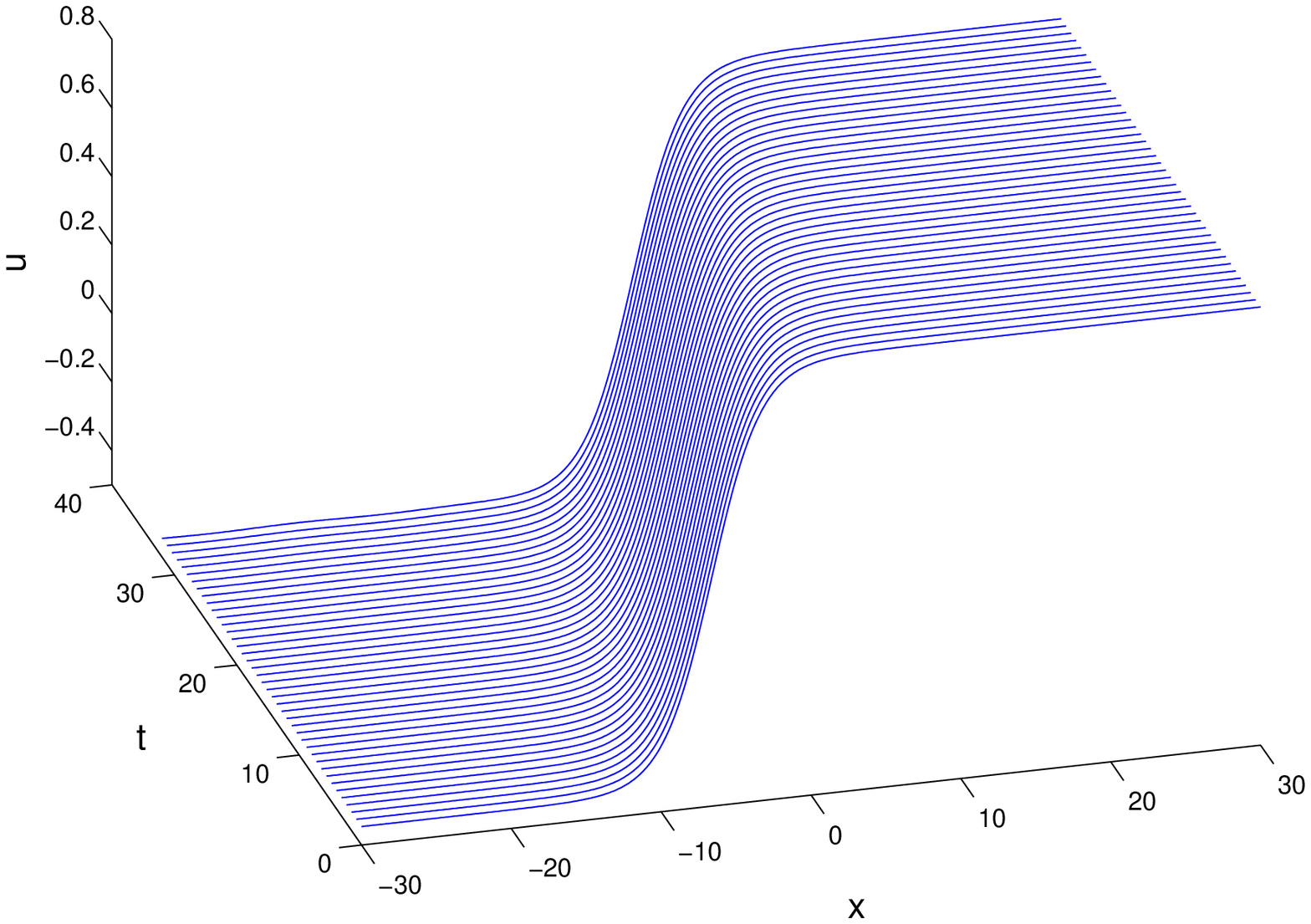}
\centerline{\hspace{0.2cm}} \caption{\label{fig:wide} (Color online)
The shock profile wave propagation of the GKS equation
(\ref{GKSE2}), $a=b=C=1, x_0 = -5$. Exact boundary condition in
$[-30, 30]$, $\Delta x=0.1, \Delta t=0.0001$.}
\end{figure*}

The next three test problems are the fifth-order Kawahara-like
equations. We use the D1Q7 LBGK model to simulate them. Since the
first six constraints on moments are enough for exactly recovering
the fifth-order NPDEs in C-E expansion, it is interesting to compare
the present \emph{exact} model with C-E expansion to order
$O(\epsilon^6)$ with one to order $O(\epsilon^5)$. We denote scheme
1 and scheme 2 for the model of order $O(\epsilon^6)$ and that of
order $O(\epsilon^5) $, respectively. For simplification, we only
take $\bar{\Pi}_6=0$ in Eq. (\ref{EDF7}) for scheme 2 in
simulations.

\emph{Example 4.5.} The Kawahara equation \cite{Wazwaz1}
\begin{equation}
u_t + \alpha u u_x +  \beta u_{xxx} + \gamma u_{xxxxx}
=0,\label{KAE}
\end{equation}
with the exact solution
\begin{eqnarray}
u(x,t)=A\mathrm{sech}^4[B(x-Ct)],
\end{eqnarray}
where $\alpha, \beta, \gamma $ are constants, and
$A=-\frac{105\beta^2}{169\alpha\gamma},B=\frac{1}{2}\sqrt{-\frac{\beta}{13\gamma}},C=-\frac{36\beta^2}{169\gamma}$.

In simulations, we set $ \alpha=\beta=-\gamma=1 $. The simulation is
conducted in $[-30,30]$ with $\Delta x=0.1, \Delta t=0.01, 0.001$
and $0.0001$. Table V. gives the errors of numerical solution for
different times. From the table
it is found that there is little difference between accuracy of two
schemes, which implies that the accuracy of higher-order model may
not be better than that of lower-order one.

\begin{table*}
\caption{Comparison of global relative errors at different times [
(1) $t=1$, (2) $t=2$; (3) $t=3$; (4) $t=4$ ].}
\begin{ruledtabular}
\begin{tabular}{ccccccccc}
  &   & & Scheme l &  &  & & Scheme 2 & \\
&$(c,\tau_{opt})$ & $(10,3.37)$ & $(100,2.53)$ & $(1000,2.02)$ & & $(10,3.35)$ & $(100,2.55)$ & $(1000,2.04)$ \\
\hline
 (1)&  &$6.0101\times 10^{-3}$&$2.6928\times 10^{-3}$&$1.5361\times 10^{-3}$&  &$5.9364\times 10^{-3}$&$2.7372\times 10^{-3}$&$1.5750\times 10^{-3}$ \\
 (2)&  &$1.0877\times 10^{-2}$&$5.2590\times 10^{-3}$&$3.0032\times 10^{-3}$&  &$1.0698\times 10^{-2}$&$5.3477\times 10^{-3}$&$3.0827\times 10^{-3}$ \\
 (3)&  &$1.5605\times 10^{-2}$&$7.5403\times 10^{-3}$&$4.2350\times 10^{-3}$&  &$1.5369\times 10^{-2}$&$7.6869\times 10^{-3}$&$4.3599\times 10^{-3}$ \\
 (4)&  &$2.0197\times 10^{-2}$&$9.4960\times 10^{-3}$&$5.3288\times 10^{-3}$&  &$2.0035\times 10^{-2}$&$9.7447\times 10^{-3}$&$5.4829\times 10^{-3}$ \\
\end{tabular}
\end{ruledtabular}
\end{table*}

\emph{Example 4.6.} The modified Kawahara equation \cite{Wazwaz2}
\begin{equation}
u_t + a u^2 u_x + b u_{xxx} - k u_{xxxxx} =0,\label{MKAE}
\end{equation}
with the exact solution
\begin{eqnarray}
u(x,t)= A\mathrm{sech}^2\left[B\left(x-Ct\right)\right],
\end{eqnarray}
where $a, b, k$ are constants, and
$A=-\frac{3b}{\sqrt{10ak}},B=\frac{1}{2}\sqrt{\frac{b}{5k}},C=\frac{4b^2}{25k}$.

In simulations, we set $a=b=k=1$. The simulation is conducted in
$[-30,30]$ with $\Delta x=0.1, \Delta t=0.01, 0.001$ and $0.0001$.
Table VI. gives the errors of numerical solution for different
times. From the table it is also found that there
is little difference between accuracy of two schemes.

\begin{table*}
\caption{Comparison of global relative errors at different times [
(1) $t=1$, (2) $t=2$; (3) $t=3$; (4) $t=4$ ].}
\begin{ruledtabular}
\begin{tabular}{ccccccccc}
  &   & & Scheme l &  &  & & Scheme 2 & \\
&$(c,\tau_{opt})$ & $(10,4.54)$ & $(100,2.53)$ & $(1000,2.04)$ & & $(10,4.04)$ & $(100,2.56)$ & $(1000,2.04)$ \\
\hline
 (1)&  &$1.9295\times 10^{-2}$&$7.1698\times 10^{-3}$&$4.4255\times 10^{-3}$&  &$1.6850\times 10^{-2}$&$7.3109\times 10^{-3}$&$4.4254\times 10^{-3}$ \\
 (2)&  &$3.8260\times 10^{-2}$&$1.3214\times 10^{-2}$&$7.7342\times 10^{-3}$&  &$3.2649\times 10^{-2}$&$1.3549\times 10^{-2}$&$7.7745\times 10^{-3}$ \\
 (3)&  &$5.7488\times 10^{-2}$&$1.7575\times 10^{-2}$&$1.0046\times 10^{-2}$&  &$4.7022\times 10^{-2}$&$1.8137\times 10^{-2}$&$1.0211\times 10^{-2}$ \\
 (4)&  &$7.4409\times 10^{-2}$&$2.1032\times 10^{-2}$&$1.1886\times 10^{-2}$&  &$6.0945\times 10^{-2}$&$2.2855\times 10^{-2}$&$1.2424\times 10^{-2}$ \\
\end{tabular}
\end{ruledtabular}
\end{table*}

\emph{Example 4.7.} The Korteweg-de Vries-Kawahara equation
\cite{Ceballos}
\begin{equation}
u_t + uu_x + u_x + u_{xxx} - u_{xxxxx} =0,\label{KdVK}
\end{equation}
with the exact solution
\begin{eqnarray}
u(x,t)=
\frac{105}{169}\mathrm{sech}^4\left(\frac{1}{2\sqrt{13}}(x-\frac{205}{169}t-x_0)\right),
\end{eqnarray}
where $x_0$ is a parameter.

In simulations, we set $x_0=20$ as in Ref. \cite{Ceballos}. The
simulation is conducted in $[0,200]$ with $\Delta x=0.1, \Delta
t=0.01, 0.001$ and $0.0001$. Table VII. gives the errors of
numerical solution for different times. It can be found that the
numerical solutions obtained by LBGK model are in good agreement with the analytic ones.
From the table it is still found that there is
little difference between accuracy of two schemes.

\begin{table*}
\caption{Comparison of global relative errors at different times [
(1) $t=1$, (2) $t=2$; (3) $t=3$; (4) $t=4$ ].}
\begin{ruledtabular}
\begin{tabular}{ccccccccc}
  &   & & Scheme l &  &  & & Scheme 2 & \\
&$(c,\tau_{opt})$ & $(10,5.01)$ & $(100,3.31)$ & $(1000,2.28)$ & & $(10,4.64)$ & $(100,2.95)$ & $(1000,2.18)$ \\
\hline
 (1)&  &$9.8169\times 10^{-3}$&$4.6248\times 10^{-3}$&$2.0662\times 10^{-3}$&  &$9.0008\times 10^{-3}$&$3.6976\times 10^{-3}$&$1.8557\times 10^{-3}$ \\
 (2)&  &$1.8335\times 10^{-2}$&$8.9484\times 10^{-3}$&$4.0584\times 10^{-3}$&  &$1.6575\times 10^{-2}$&$7.1666\times 10^{-3}$&$3.6371\times 10^{-3}$ \\
 (3)&  &$2.6841\times 10^{-2}$&$1.3201\times 10^{-2}$&$6.1428\times 10^{-3}$&  &$2.3973\times 10^{-2}$&$1.0468\times 10^{-2}$&$5.2117\times 10^{-3}$ \\
 (4)&  &$3.5872\times 10^{-2}$&$1.8257\times 10^{-2}$&$7.9304\times 10^{-3}$&  &$3.1522\times 10^{-2}$&$1.4085\times 10^{-2}$&$6.6671\times 10^{-3}$ \\
\end{tabular}
\end{ruledtabular}
\end{table*}

The last two test problems are the third-order KdV-type equations.
We use the D1Q5 LBGK model to simulate them. Similar to the above
three test problems, we also compare the present \emph{exact} model
with C-E expansion to order $O(\epsilon^4)$ with one to order
$O(\epsilon^3)$, and denote scheme 1 and scheme 2 for the model of
order $O(\epsilon^4)$ and that of order $O(\epsilon^3)$,
respectively. For simplification, we only take $\bar{\Pi}_4=0$ in
Eq. (\ref{EDF5}) for scheme 2 in simulations.

\emph{Example 4.8.} The KdV Burgers equation \cite{ChaiShiZheng}
\begin{equation}
u_t + \alpha uu_x - \gamma u_{xx}+ \delta u_{xxx} =0,\label{KdVB}
\end{equation}
with the exact solution
\begin{eqnarray}
u(x,t)= 2\xi \left(1-\frac{1}{[1+e^{2\nu(x-\xi t)}]^{2}}\right),
\end{eqnarray}
where $\nu=-\frac{\gamma}{10\delta}$ and
$\xi=\frac{6\gamma^2}{25\delta}$.

In simulations, we set $\alpha=1,\gamma=9\times
10^{-4},\delta=2\times 10^{-5}$, and the simulation is conducted in
$[-4,4]$ with $\Delta x=\Delta t=0.01$ as in Refs.
\cite{ChaiShiZheng} and \cite{MaCF1}. Table VIII. gives the errors
of numerical solution for different times. From the table it is
found that the errors of schemes 1 and 2 are much smaller than those
of the model in Ref. \cite{MaCF1}, and scheme 1 is better that
scheme 2, but there is little difference between our schemes.
The numerical results show that the model in Ref. \cite{MaCF1} is not exact, even for the
constraints on lower-order moments.

\begin{table*}
\caption{Comparison of global relative errors for $c=1$ at different
times [ (1) Scheme 1; (2) Scheme 2; (3) Scheme in Ref. \cite{MaCF1}
].}
\begin{ruledtabular}
\begin{tabular}{cccccccc}

  $\tau_{opt}$ &    &$t=10$&$t=50$&$t=150$&$t=250$&$t=300$ \\
\hline
 $0.97$& (1) &$2.4300\times 10^{-6}$&$4.2738\times 10^{-6}$&$4.0518\times 10^{-6}$&$3.4676\times 10^{-6}$&$3.3242\times 10^{-6}$\\
 $0.96$& (2) &$4.9172\times 10^{-6}$&$7.1775\times 10^{-6}$&$6.1615\times 10^{-6}$&$5.2459\times 10^{-6}$&$4.9383\times 10^{-6}$ \\
 $0.968$& (3) &$1.0416\times 10^{-5}$&$1.8801\times 10^{-5}$&$1.7409\times 10^{-5}$&$1.4877\times 10^{-5}$&$1.3901\times 10^{-5}$ \\

\end{tabular}
\end{ruledtabular}
\end{table*}

\emph{Example 4.9.} The K($n,n$)-Burgers equation \cite{Wazwaz3}
\begin{equation}
u_t + a (u^n)_x + b (u^n)_{xxx} +k u_{xx} =0,
\end{equation}
with the exact solution
\begin{eqnarray}
u(x,t)= \left[A(1+\tanh(Bx+Ct))\right]^{-\frac{1}{n-1}}, b>0,n>1,a<0
\end{eqnarray}
where $a,b,k,n$ are constants, and $A=\frac{1}{2k}\sqrt{-ab}, B=-\frac{n-1}{2n}\sqrt{-\frac{a}{b}}, C=\frac{ak(n-1)}{2bn}$.

In simulations, we set $a=-1,b=1,k=-1,n=2$. The simulation is
conducted in $[-1,1]$ with $\Delta x=0.01, \Delta t=0.001$. Table
IX. gives the errors of numerical solution for different times. From
the table it is found that the errors of schemes 1 are much smaller
than those of scheme 2, which implies that for this problem the
accuracy of higher-order model is much better than that of
lower-order one.

\begin{table*}
\caption{Comparison of global relative errors for $c=10$ at
different times.}
\begin{ruledtabular}
\begin{tabular}{ccccccc}

 $\tau_{opt}$  &    &$t=1$&$t=2$&$t=3$&$t=4$ \\
\hline
 $37.77$& Scheme 1 &$1.8629\times 10^{-3}$&$9.1678\times 10^{-4}$&$9.2998\times 10^{-4}$&$7.1608\times 10^{-4}$\\
 $31.52$& Scheme 2 &$2.6472\times 10^{-2}$&$1.0947\times 10^{-2}$&$6.1720\times 10^{-3}$&$3.3151\times 10^{-3}$ \\

\end{tabular}
\end{ruledtabular}
\end{table*}

\section{Conclusion}

In the present work, we have developed a unified LBGK model for 1D
higher-order NPDEs. Through C-E expansion a given NPDE can be exactly recovered
 to required order of small parameter $\epsilon$ by choosing correct auxiliary moments.
 Unlike traditional numerical methods which solve
for macroscopic variables, the model has the advantages of standard
LBGK model, which are borrowed from kinetic theory, such as
linearity of the convection operator in velocity space, simplicity
and symmetry of scheme, ease in coding and intrinsical parallelism
\cite{cd}. Detailed numerical tests of the proposed model are
carried out for different types of NPDEs, including the
Kuramoto-Sivashinsky type equations, Kawahara type equations, and
KdV type equations. It is found that the simulation results agree
well with the analytical and numerical solutions reported in
previous studies, which shows that the LBM has potentials in
simulating higher-order NPDEs. However, perhaps due to the effects
of nonlinearity and higher order differentials, the LBGK model for
solving higher order NPDEs is sensitive to the key parameters, such
as $\Delta x, \Delta t$ and $\tau$, and it does not seems so
efficient as that for solving lower order ones, such as that for
NCDEs \cite{ShiGuo}.

Note that the proposed model can be directly applied to derive the
LBGK model for high-order NPDEs in higher dimensional space by
treating moments as tensors, and the LBGK model for NPDEs with order larger
than six can be easily derived by using the idea in this paper.
Nevertheless, some important issues,
such as how to improve the accuracy and stability of the LB models
need further studies.

\begin{acknowledgments}
This work is supported by the National Science
Foundation of China (Grants No. 60773195 and No. 50606012).
\end{acknowledgments}

\end{document}